\documentclass[longauth]{aa} 

\usepackage{graphicx}
\usepackage{txfonts}

\usepackage{subfigure}
\usepackage{natbib}
\usepackage{amssymb,amsmath}
\usepackage{array}
\usepackage{booktabs}

\usepackage{mhchem}
\usepackage{threeparttable}
\usepackage{sidecap}
\sidecaptionvpos{figure}{t}

\usepackage[colorlinks=true,urlcolor=blue,linkcolor=blue,citecolor=blue]{hyperref}

\begin{document}

    \title{TOI-2257\,b: A highly eccentric long-period sub-Neptune transiting a nearby M dwarf}
    \titlerunning{TOI-2257\,b}

    \subtitle{}

    \author{N.~Schanche \inst{\ref{unibe}} 
        \and F.J.~Pozuelos\inst{\ref{aru_liege},\ref{star_liege}}
        \and M.N.~G\"unther\inst{\ref{mitkavli},\ref{estec}}$^,$\thanks{Juan Carlos Torres Fellow}$^,$\thanks{ESA Research Fellow}
        \and R.D.~Wells\inst{\ref{unibe}}
        \and A.J.~Burgasser\inst{\ref{ucsd}}
        \and P.~Chinchilla\inst{\ref{aru_liege},\ref{iac}}
        \and L.~Delrez\inst{\ref{aru_liege},\ref{star_liege}}
        \and E.~Ducrot\inst{\ref{aru_liege}}
        \and L.J.~Garcia\inst{\ref{aru_liege}}
        \and Y.~G\'omez~Maqueo~Chew\inst{\ref{ciudad}}
        \and E.~Jofr\'e\inst{\ref{ciudad}, \ref{cordoba}, \ref{conicet}}
        \and B.V.~Rackham\inst{\ref{miteaps},\ref{mitkavli}}$^,$\thanks{51 Pegasi b Fellow}
        \and D.~Sebastian\inst{\ref{birmingham}}
        \and K.G.~Stassun\inst{\ref{vanderbilt}}
        \and D.~Stern\inst{\ref{jpl}}
        \and M.~Timmermans\inst{\ref{aru_liege}}
        \and K.~Barkaoui\inst{\ref{aru_liege},\ref{oukaimeden}}
        \and A.~Belinski\inst{\ref{sternberg}}
        \and Z.~Benkhaldoun\inst{\ref{oukaimeden}}
        \and W.~Benz\inst{\ref{unibe}, \ref{BernPhysics}}
        \and A.~Bieryla\inst{\ref{cfa}}
        \and F.~Bouchy\inst{\ref{geneva}}
        \and A.~Burdanov\inst{\ref{miteaps}}
        \and D.~Charbonneau\inst{\ref{cfa}}
        \and Jessie L.~Christiansen\inst{\ref{caltech}}
        \and Karen A.\ Collins\inst{\ref{cfa}}
        \and B.-O.~Demory\inst{\ref{unibe}}
        \and M.~D\'evora-Pajares\inst{\ref{granada}}
        \and J.~de~Wit\inst{\ref{miteaps}}
        \and D.~Dragomir\inst{\ref{unm}}
        \and G.~Dransfield\inst{\ref{birmingham}}
        \and E.~Furlan\inst{\ref{nesi}}
        \and M.~Ghachoui\inst{\ref{aru_liege},\ref{oukaimeden}}
        \and M.~Gillon\inst{\ref{aru_liege}}
        \and C.~Gnilka\inst{\ref{ames}}
        \and M.A.~G\'omez-Mu\~noz\inst{\ref{uname}}
        \and N.~Guerrero\inst{\ref{florida},\ref{mitkavli}}
        \and M.~Harris\inst{\ref{unm}}
        \and K.~Heng\inst{\ref{unibe}, \ref{warwick}}
        \and C.E.~Henze\inst{\ref{ames}}
        \and K.~Hesse\inst{\ref{mitkavli}}
        \and S.B.~Howell\inst{\ref{ames}}
        \and E.~Jehin\inst{\ref{star_liege}}
        \and J.~Jenkins\inst{\ref{ames}}
        \and Eric L.\ N.\ Jensen\inst{\ref{swarthmore}}
        \and M.~Kunimoto\inst{\ref{mitkavli}}
        \and D.W.~Latham\inst{\ref{cfa}}
        \and K.~Lester\inst{\ref{ames}}
        \and Kim K. McLeod\inst{\ref{wellesley}}
        \and I.~Mireles\inst{\ref{unm}}
        \and C.A.~Murray\inst{\ref{cavendish}}
        \and P.~Niraula\inst{\ref{miteaps}}
        \and P.P.~Pedersen\inst{\ref{cavendish}}
        \and D.~Queloz\inst{\ref{cavendish}}
        \and E.~V.~Quintana\inst{\ref{goddard}}
        \and G.~Ricker\inst{\ref{mitkavli}}
        \and A.~Rudat\inst{\ref{mitkavli}}
        \and L.~Sabin\inst{\ref{uname}}
        \and B.~Safonov\inst{\ref{sternberg}}
        \and U.~Schroffenegger\inst{\ref{unibe}}
        \and N.Scott\inst{\ref{ames}}
        \and S.~Seager\inst{\ref{mitkavli},\ref{miteaps},\ref{mitaeroastro}}
        \and I.~Strakhov\inst{\ref{sternberg}}
        \and A.H.M.J.~Triaud\inst{\ref{birmingham}}
        \and R.~Vanderspek\inst{\ref{mitkavli}}
        \and M.~Vezie\inst{\ref{mitkavli}}
        \and J.~Winn\inst{\ref{princeton}}
        }

\institute{
    Center for Space and Habitability, University of Bern, Gesellschaftsstrasse 6, 3012, Bern, Switzerland \label{unibe}
    \and Astrobiology Research Unit, Universit\'e de Li\`ege, All\'ee du 6 Ao\^ut 19C, B-4000 Li\`ege, Belgium \label{aru_liege}
    \and Space Sciences, Technologies and Astrophysics Research (STAR) Institute, Universit\' de Li\`ege, All\'ee du 6 Ao\^ut 19C, B-4000 Li\`ege, Belgium \label{star_liege}
    \and Department of Physics, and Kavli Institute for Astrophysics and Space Research, Massachusetts Institute of Technology (MIT), Cambridge, MA 02139, USA \label{mitkavli}
    \and European Space Agency (ESA), European Space Research and Technology Centre (ESTEC), Keplerlaan 1, 2201 AZ Noordwijk, The Netherlands \label{estec}
    \and Center for Astrophysics and Space Science, University of California San Diego, La Jolla, CA 92093, USA \label{ucsd}
    \and Instituto de Astrof\'isica de Canarias (IAC), Calle V\'ia L\'actea s/n, 38200, La Laguna, Tenerife, Spain \label{iac}
    \and Universidad Nacional Aut\'onoma de M\'exico, Instituto de Astronom\'ia, AP 70-264, CDMX  04510, M\'exico \label{ciudad}
    \and Universidad Nacional de C\'ordoba - Observatorio Astron\'omico de C\'ordoba, Laprida 854, X5000BGR, C\'ordoba, Argentina \label{cordoba}
    \and Consejo Nacional de Investigaciones Cient\'ificas y T\'ecnicas (CONICET), Argentina \label{conicet}
    \and Department of Earth, Atmospheric and Planetary Science, Massachusetts Institute of Technology, 77 Massachusetts Avenue, Cambridge, MA 02139, USA \label{miteaps}
    \and School of Physics \& Astronomy, University of Birmingham, Edgbaston, Birmimgham B15 2TT, UK \label{birmingham}
    \and Department of Physics \& Astronomy, Vanderbilt University, 6301 Stevenson Center Ln., Nashville, TN 37235, USA \label{vanderbilt}
    \and Jet Propulsion Laboratory, California Institute of Technology, 4800 Oak Grove Drive, MS 169-224, Pasadena, CA 91109, USA \label{jpl}
    \and Oukaimeden Observatory, High Energy Physics and Astrophysics Laboratory, Cadi Ayyad University, Marrakech, Morocco \label{oukaimeden}
    \and Sternberg Astronomical Institute, Moscow State University (SAI MSU), Universitetskii pr. 13, Moscow, 119991 Russia \label{sternberg}
    \and Physikalisches  Institut,  University  of  Bern,  Gesellsschaftstrasse  6,3012 Bern, Switzerland \label{BernPhysics}
    \and Center for Astrophysics | Harvard \& Smithsonian, 60 Garden Street, Cambridge, MA, 02138, USA \label{cfa}
    \and Observatoire de l’Universit\'e de Gen\'eve, Chemin des Maillettes 51, CH-1290 Versoix, Switzerland \label{geneva}
    \and Caltech/IPAC, 1200 E. California Boulevard, Pasadena, CA 91125, USA \label{caltech}
    \and Dpto. F\'isica Te\'orica y del Cosmos, Universidad de Granada, 18071, Granada, Spain \label{granada}
    \and Department of Physics and Astronomy, University of New Mexico, 1919 Lomas Blvd NE, Albuquerque, NM 87131, USA \label{unm}
    \and NASA Exoplanet Science Institute, Caltech/IPAC, Mail Code 100-22, 1200 E. California Blvd., Pasadena, CA 91125, USA \label{nesi}
    \and NASA Ames Research Center, Moffett Field, CA 94035, USA \label{ames}
    \and Universidad Nacional Aut\'onoma de M\'exico, Instituto de Astronom\'ia, AP 106, Ensenada 22800, BC, M\'exico \label{uname}
    \and Department of Astronomy, University of Florida, Gainesville, FL, 32611, USA \label{florida}
    \and Department of Physics, University of Warwick, Gibbet Hill Road, Coventry CV4 7AL, United Kingdom \label{warwick}
    \and Dept. of Physics \& Astronomy, Swarthmore College, Swarthmore PA 19081, USA \label{swarthmore}
    \and Department of Astronomy, Wellesley College, Wellesley, MA 02481, USA \label{wellesley}
    \and Cavendish Laboratory, JJ Thomson Avenue, Cambridge, CB3 0HE, UK \label{cavendish}
    \and NASA Goddard Space Flight Center, 8800 Greenbelt Road, Greenbelt, MD 20771, USA \label{goddard}
    \and Department of Aeronautics and Astronautics, MIT, 77 Massachusetts Avenue, Cambridge, MA 02139, USA \label{mitaeroastro}
    \and Department of Astrophysical Sciences, Princeton University, 4 Ivy Lane, Princeton, NJ 08544, USA \label{princeton}
}


    \date{Received ; accepted }

\abstract
    {Thanks to the relative ease of finding and characterizing small planets around M dwarf stars, these objects have become cornerstones in the field of exoplanet studies. The current paucity of planets in long-period orbits around M dwarfs make such objects particularly compelling as they provide clues about the formation and evolution of these systems. } 
    {In this study, we present the discovery of TOI-2257\,b (TIC 198485881), a long-period (35\,d) sub-Neptune orbiting an M3 star at 57.8\,pc. Its transit depth is about 0.4\,\%, large enough to be detected with medium-size, ground-based telescopes. The long transit duration suggests the planet is in a highly eccentric orbit ($e \sim 0.5$), which would make it the most eccentric planet that is known to be transiting an M-dwarf star.}
    {We combined TESS and ground-based data obtained with the 1.0-m SAINT-EX, 0.60-m TRAPPIST-North and 1.2-m FLWO telescopes to find a planetary size of 2.2\,$R_{\oplus}$ and an orbital period of 35.19\,days. In addition, we make use of archival data, high-resolution imaging, and vetting packages to support our planetary interpretation. }
    {With its long period and high eccentricity, TOI-2257\,b falls in a novel slice of parameter space. Despite the planet's low equilibrium temperature ($\sim$\,256 K), its host star's small size ($R_* = 0.311 \pm{0.015}$) and relative infrared brightness (K$_{mag}$ = 10.7) make it a suitable candidate for atmospheric exploration via transmission spectroscopy.}
    {}

\keywords{Planets and satellites: detection -- Stars: individual: TOI-2257 -- Stars: individual: TIC 198485881 -- Techniques: photometric}

\maketitle


\section{Introduction}
Despite their cool temperatures, M dwarfs have become hot targets for exoplanet surveys. The large planet-to-star ratios of these systems result in relatively deep transits and large radial velocity amplitudes. In addition, they emit more strongly in the infrared. This leads to favorable conditions for atmospheric characterization by the Hubble Space Telescope (HST) and James Webb Space Telescope (JWST). Several ground-based surveys, such as SPECULOOS \citep{delrez_speculoos_2018, Sebastian2021} and MEarth \citep{Mearth} via the transit method and CARMENES \citep{CARMENES} and the M-dwarf sample of HARPS \citep{Bonfils2013} using radial velocities, have been created with the specific goal of finding planets around M dwarfs. 

While space-based discoveries have been led by \textit{Kepler} \citep{Kepler} and K2 \citep{K2}, the Transiting Exoplanet Survey Satellite \citep[TESS;][]{TESS} is making a growing contribution to the population of planet-hosting M dwarfs, with 28 of 185 such planets discovered. Of those, only 17 are transiting planets in long-period (>15 day) orbits\footnote{\url{https://exoplanetarchive.ipac.caltech.edu/}}. 

In this paper, we present the discovery and statistical validation of a long-period sub-Neptune orbiting an M3 star. Intriguingly, the planet's orbit is highly elliptical, suggestive of a possible perturbing outer gas giant \citep{Huang2017,VanEylen2019}. The planet was first identified in TESS data, with further confirmation from several ground-based facilities, including three telescopes in the SPECULOOS group \citep{Sebastian2021, murray_photometry_2020}, which are designed for observations of small planets around ultra-cool M-dwarfs. The planet is further validated with contributions from high-resolution imaging, spectroscopy, and archival data.

The structure of the present work is as follows. In Section \ref{sec:obs} we introduce the contributing facilities and datasets used in the validation of TOI-2257\,b. Section \ref{sec:stellar} describes the work to characterize the host star. The analysis of the transit lightcurves and characterization of the system are presented in Section \ref{sec:fit}. In Section \ref{sec:validation} we discuss possible false positive scenarios and argue that there is sufficient evidence that the transits are validated as planetary in nature. The possibility of an additional planetary companion in the system is explored in Section \ref{sec:search}. Finally, implications for this planet are discussed in Section \ref{sec:discussion}.


\section{Observations} \label{sec:obs}
In this section, we present all observations of TOI-2257 taken by TESS and ground-based follow-up facilities. A summary of the photometric follow-up observations is shown in Table \ref{tab:follow_up_obs}.

\subsection{TESS photometry} 
TOI-2257 is a part of the TESS Candidate Target List \citep{Stassun2018} and was observed with a 2-min cadence in TESS sectors 14 (18 July 2019-16 Aug 2019), 20 (24 Dec 2019-21 Jan 2020), 21 (21 Jan 2020-18 Feb 2020), and 26 (8 Jun 2020-4 Jul 2020). The image data were processed by the Science Processing Operations Center (SPOC) pipeline \citep{Jenkins2016} to produce photometric time series which were searched for transiting planet signatures with a noise-compensating matched filter \citep{Jenkins2002,Jenkins2010, Jenkins2020} which detected a pair of transits separated by 175 days. The transit signature passed all the diagnostic tests performed on the data \citep{Twicken2018,Li2019}, including the difference image centroid test, which located the source of the transit signature to within 1.5 $\pm$ 4.4 arcsec. No additional transit signatures were found in the subsequent multiple planet search. The TESS Science Office reviewed the data validation reports and issued an alert for this planet candidate on 10 August 2020 \citep{Guerrero2021}.

\begin{SCfigure*}
    \centering
    \includegraphics[width=0.75\textwidth]{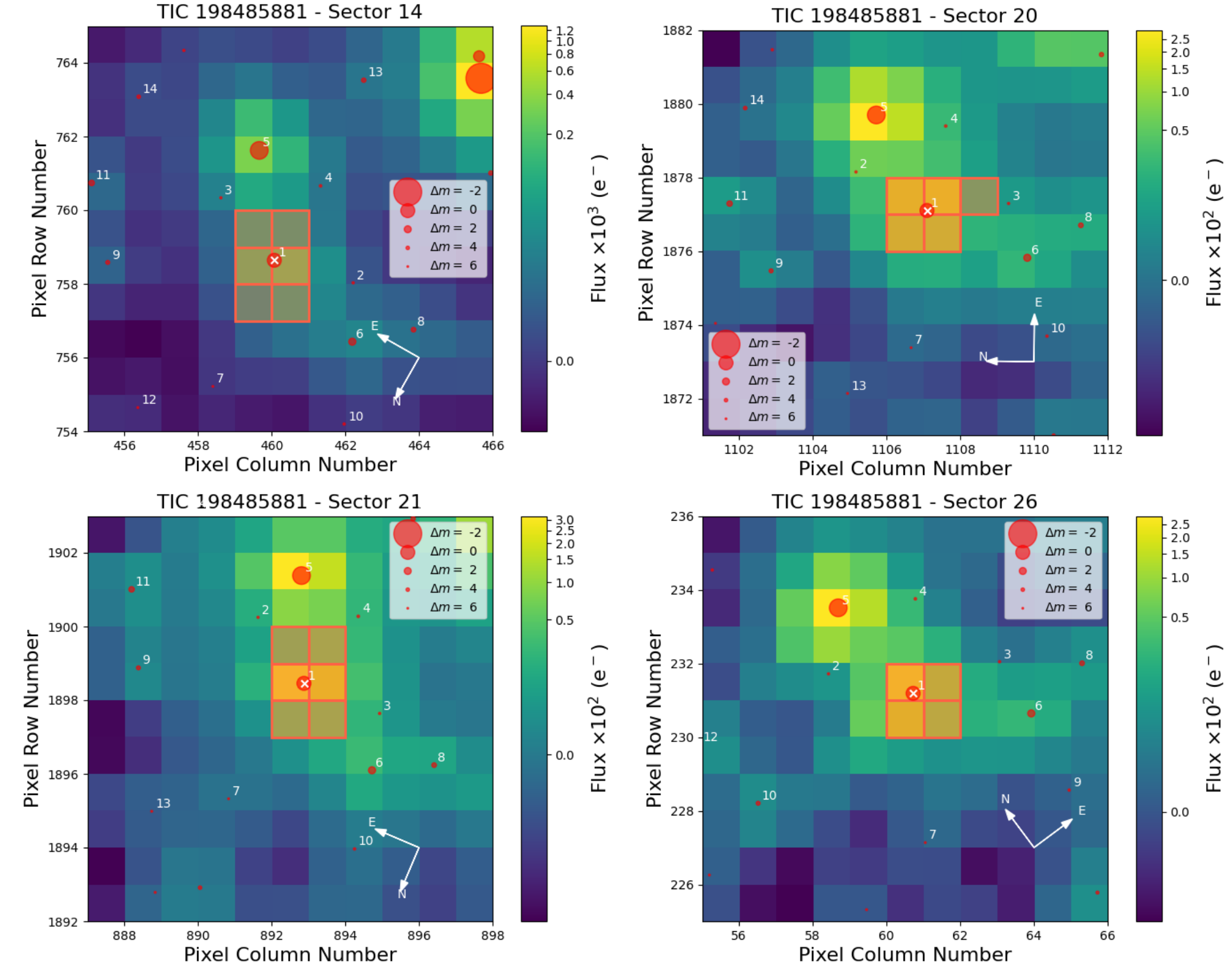}
    \caption{\textit{TESS} target pixel files (TPFs) of Sectors 14, 20, 21 and 26 that observed TOI-2257, generated by means of \texttt{tpfplotter} (\citealp{aller:2020}). The apertures used to extract the photometry by the SPOC pipeline are shown as red shaded regions. The {\it Gaia\/} DR2 catalog \citep{gaia18} is over-plotted, with all sources up to 6 magnitudes in contrast with TOI-2257 shown as red circles. We note that the symbol size scales with the magnitude contrast. While the star is relatively isolated, there is a small amount of contamination from outside sources, ranging from 2-5\% of the total flux.}
    \label{fig:tpf}
\end{SCfigure*}

\begin{figure*}
    \centering
    \includegraphics[width=0.999\textwidth]{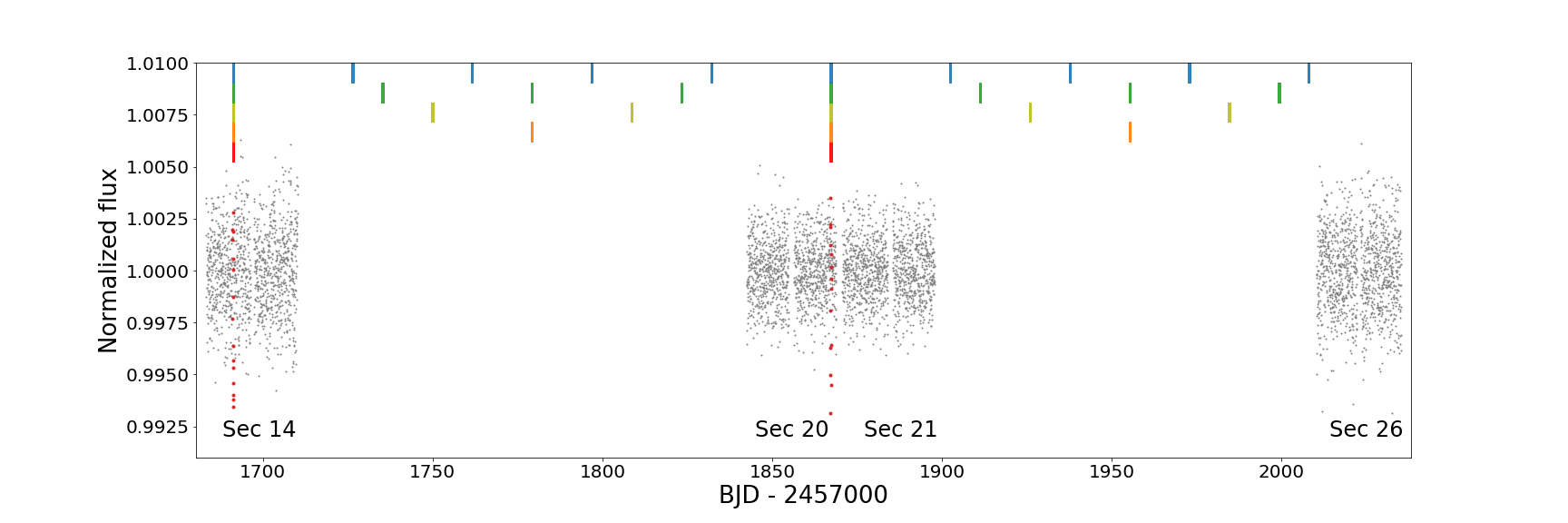}
    \caption{Possible orbital solutions for the observed TESS transits. In-transit data points are shown in red. Based on the pattern of detections and the observation spacing, possible orbital periods include 176, 88, 59, 44, and 35 days, shown by the red, orange, mustard, green, and blue ticks respectively. }
    \label{fig:TESS_aliases}
\end{figure*}

Across the four sectors, only two transit events were detected. The separation of the two observed transits and the gaps in coverage of the object led to a possible period of 175.9 days but could not rule out other possibilities including 88\,d, 58.6\,d, 44\,d, and 35.2\,d (See Fig. \ref{fig:TESS_aliases}).
From the Mikulski Archive for Space Telescopes (MAST) database, we obtained the Presearch Data Conditioning Simple Aperture Photometry \citep[PDC-SAP;][]{Smith2012,Stumpe2012,Stumpe2014}, in which long-term trends in the data are removed. In addition, we remove all datapoints for which the quality flag was not 0. Fig.~\ref{fig:tpf} shows the apertures used for the TESS data with Gaia DR2 sources overlaid on the images.

\subsection{Follow-up photometry} 
Several ground-based observations of TOI-2257 were taken in order to secure the period derived from TESS, constrain the transit parameters, and characterize the star. A non-detection on 4 March 2021 using the LCO McDonald 1-m telescope ruled out the 58-day alias. The 35-day alias was sampled next on 20 Apr 2021, producing a clear transit event on three separate telescopes, securing that period solution. A further full transit was obtained during the next transit window. We describe each of these observations in the following section. The data themselves will be shown in Section \ref{sec:fit}.

\begin{table*}[]
    \centering
    \caption{Ground-based photometric time series observations of TOI-2257.}
    \begin{tabular}{l c c c c}
    \hline
    \hline
         Date (UT) & Filter & Facility & Exp. time (s) & Notes \\
         \hline
         20 Apr 2021 & I+z & SAINT-EX & 20 & partial (egress) \\
         20 Apr 2021 & $i'$ & KeplerCam & 60 & partial (egress) \\
         20 Apr 2021 & $z'$ & TRAPPIST-N & 60 & partial (ingress) \\
         25 May 2021 & I+z & SAINT-EX & 25 & full \\
         25 May 2021 & $r'$ & KeplerCam & 120 & partial (ingress) \\
         25 May 2021 & $z'$ & KeplerCam & 90 & partial (ingress) \\
        
         \hline
    \end{tabular}

    \label{tab:follow_up_obs}
\end{table*}

\subsubsection{LCOGT}\label{sec:lcogt}

We observed TOI-2257 in Sloan $i'$ UTC 2021 March 04 from the Las Cumbres Observatory Global Telescope (LCOGT) \citep{Brown:2013} 1.0\,m network node at McDonald Observatory. We used the {\tt TESS Transit Finder}, which is a customized version of the {\tt Tapir} software package \citep{Jensen2013}, to schedule our transit observation. The $4096\times4096$ LCOGT SINISTRO cameras have an image scale of $0\farcs389$ per pixel, resulting in a $26\arcmin\times26\arcmin$ field of view. The images were calibrated by the standard LCOGT {\tt BANZAI} pipeline \citep{McCully:2018}, and photometric data were extracted with {\tt AstroImageJ} \citep{Collins2017}. The observation covered the 58.6 day orbital period alias. The expected transit was ruled out, thus eliminating this orbital period from the list of aliases allowed by the TESS data.

\subsubsection{TRAPPIST-North}
We observed TOI-2257 on UTC 2021 April 20 from the TRAPPIST-North telescope, located at the Ouka\"{i}meden observatory in Morocco \citep{jehin2011, gillon2011, barkaoui2019}. TRAPPIST-North is a 0.6-m Ritchey-Chr\'etien telescope equipped with an ANDOR CCD camera providing a field of view of $20'\times20'$ with 0.60\arcsec pixels. This observation consisted of 369 frames of 60s exposure in z' filter, and captured the ingress of TOI-2257 planetary candidate transit. The data were reduced using \texttt{prose} \citep{prose_paper}, including frame alignment, calibration, and photometric extraction.

\subsubsection{SAINT-EX}
\label{saintex}
We used the Search And characterIsatioN of Transiting EXoplanets (SAINT-EX) Observatory \citep{demory:2020} to obtain one partial and one full transit of TOI-2257. SAINT-EX is a 1-m F/8 Ritchey-Chr\'etien telescope located at the Observatorio Astron\'omico Nacional in the Sierra de San Pedro M\'artir, M\'exico. SAINT-EX has an Andor iKon-L camera with deep-depletion e2v $2k\times2k$ CCD optimized in the near infrared (NIR). The detector field of view is $12'\times12'$ with 0.34" per pixel.

The observations were made with an I+z filter (transmittance $>$90\% from 750\,nm to beyond 1000\,nm) designed for the observation of faint red targets usually observed by the SPECULOOS survey \citep{delrez_speculoos_2018,murray_photometry_2020,Sebastian2021}. The first observation sequence on 20 April 2021 contained 511 observations with 20s cadence and began shortly after the expected ingress. A full transit was observed 35 days later during the next transit window on May 25, 2021 with 402 observations using 25s cadence. Due to the low viewing angle, the telescope was re-oriented shortly before egress, resulting in a slight reduction in precision in the final 84 images. The data was reduced using the custom pipeline PRINCE \citep{demory:2020}.

\subsubsection{FLWO}

We observed TOI-2257 on UTC 2021 April 20 from the Fred Lawrence Whipple Observatory (FLWO) on Mt. Hopkins, Arizona, USA, using the KeplerCam CCD on the 1.2-m telescope. KeplerCam is a $4096\times4096$ detector used in $2\times2$ binning mode, producing a $0\farcs672$ per pixel scale and a $23.1\arcmin\times23.1\arcmin$ field of view. The images were reduced using standard IDL routines, and AstroImageJ was used to perform aperture photometry. We obtained 245 observations in the $i'$ band over a period of 302 minutes, resulting in a clear egress detection on target.

Further observations were taken on UTC 2021 May 25 using the KeplerCam CCD in alternating $z'$ and $r'$ filters for a total of 44 measurements in each filter with exposure times of 90 and 120 seconds, respectively. 

\subsubsection{SPECULOOS North}
In addition to observations of the transit events, we also obtain 93.5 hours of observations to monitor TOI-2257 from the telescopes associated with the SPECULOOS group and located in the Northern Hemisphere, i.e., the SAINT-EX telescope as described in Section \ref{saintex} and Artemis at the SPECULOOS Northern Observatory (SNO) at Teide Observatory in Tenerife \citep{delrez_speculoos_2018}. SNO data were reduced using the \texttt{prose} \citep{prose_paper} Python framework and SAINT-EX data were reduced using the PRINCE (Photometric Reduction and In-depth Nightly Curve Exploration) custom pipeline. The results of this monitoring are provided in Sec \ref{sec:speculoos_photometry}.

\subsection{Spectroscopy}
In order to better constrain the stellar properties, we also obtained two spectra. The results of the spectral analysis are included in Section \ref{sec:stellar}.
\subsubsection{NOT/ALFOSC}

TOI-2257 was observed with the Alhambra Faint Object Spectrograph and Camera (ALFOSC) mounted on the 2.5 m Nordic Optical Telescope (NOT) at the Observatorio Roque de los Muchachos (ORM) in La Palma, on 17 June 2021, in long slit spectroscopy mode. The grism used was gr5, covering the wavelength range between 5000–10\,700 \AA, and the slit width was 1\farcs3, leading to a resolution of $\sim$320. Two exposures of 90\,s were obtained. The spectroscopic standard star SP1446+259 was also observed to correct for instrumental response. Some bias, flat field and arc images were also acquired for the reduction and calibration of the spectra.

The reduction was performed using IRAF standard routines, which included bias subtraction, flat field correction, wavelength calibration and extraction of the spectra. The instrumental response was corrected using the spectrum of the spectroscopic standard star. The spectra were not corrected for atmospheric telluric absorptions.

\subsubsection{DBSP}

We observed TOI-2257 on UTC 2021 June 16 from the Hale 200'' telescope at Palomar Observatory using the dual-beam, optical Double Spectrograph (DBSP). The night suffered from variable cloud cover, and seeing at the start of the night was $\sim 1\farcs5 - 2\farcs0$. We obtained two 450~s observations using the 1\farcs5 slit, the 5600\AA\ dichroic, the 600~$\ell$~mm$^{-1}$ blue grism ($\lambda_{\rm blaze} = 4000$~\AA), and the 400~$\ell$~mm$^{-1}$ red grating ($\lambda_{\rm blaze} = 8500$~\AA). This instrument configuration covers the full optical window at moderate resolving power, $R \approx 1000$, with a modest gap at the dichroic. Relative flux calibration was obtained using observations of the subdwarf O-star HZ44 from Massey \& Gronwall (1990; ApJ, 358, 344) obtained on the same night. The spectra were reduced using standard IRAF routines.


\subsection{High-resolution imaging} \label{sec:HiRes}

Two high angular resolution images were taken as part of the TFOP in order to rule out false positives caused by possible unresolved stellar companions or detect any close companions that could dilute the transit signal, leading to an underestimation of the planet radius. These observations are presented below. 

\subsubsection{SPP Speckle Interferometry}

\begin{figure}
    \centering
    \includegraphics[width=0.48\textwidth]{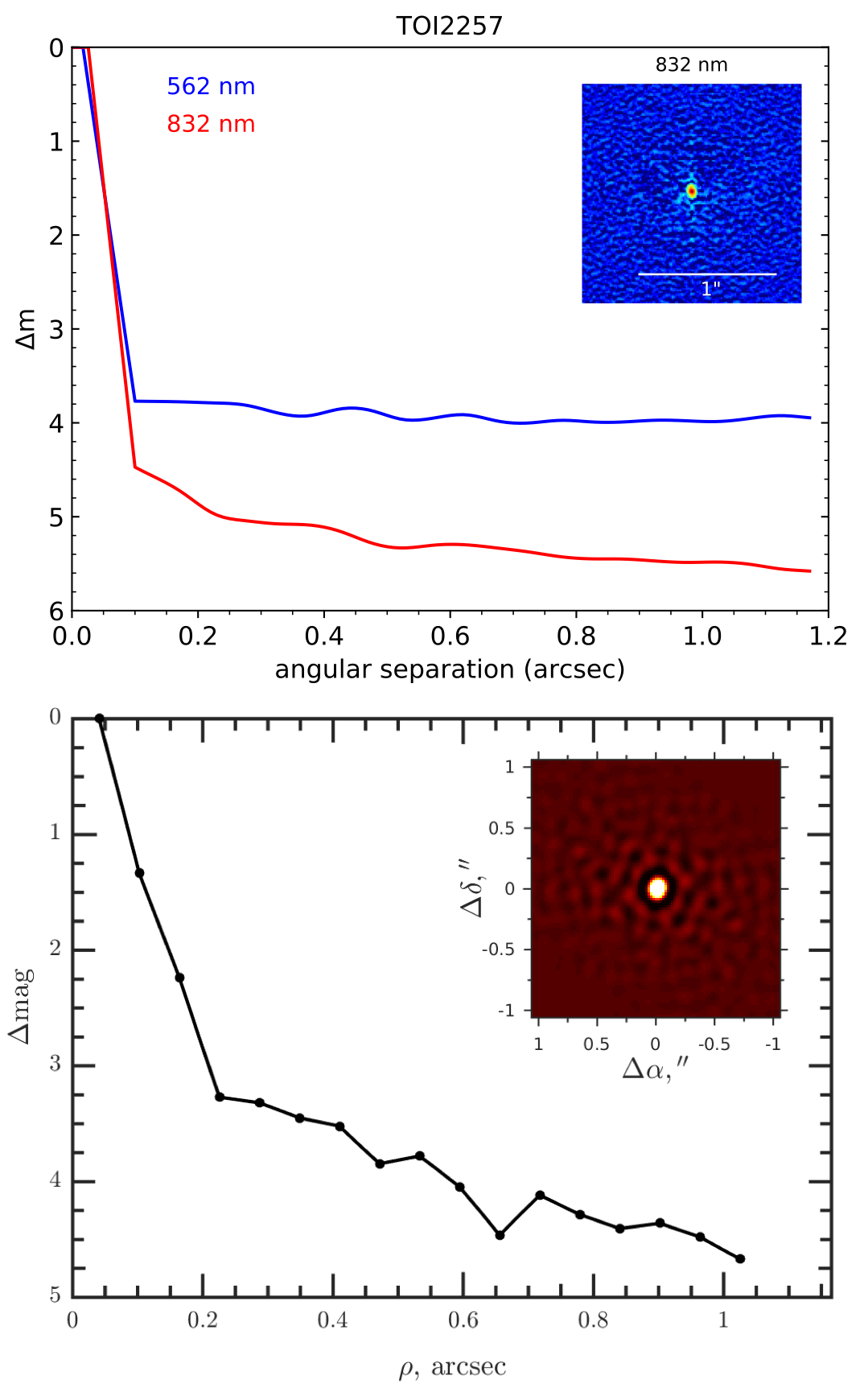}
    \caption{Contrast curves for TOI-2257\,b from high angular resolution imaging. The top plot shows the results of the Alopeke speckle instrument on Gemini North. The bottom plot shows the speckle interferometry taken by the SPeckle Polarimeter at the Sternberg Astronomical Institute. Both plots have the final reconstructed image inlaid in the upper right. Neither instrument detects a nearby companion with a magnitude within 4.5 mags of the target star.}
    \label{fig:hires}
\end{figure}

TOI-2257 was observed using speckle interferometry on 2020 November 29 with the SPeckle Polarimeter (SPP; \citealt{Safonov2017}) on the 2.5-m telescope at the Sternberg Astronomical Institute of Lomonosov Moscow State University (SAI MSU). SPP uses Electron Multiplying CCD Andor iXon 897 as a detector with a pixel scale of 20.6~mas px$^{-1}$. The observation was made in the $I$ band, and the atmospheric dispersion was compensated. The angular resolution was 89~mas. The detection limit for faint stellar companions is provided in Figure~\ref{fig:hires}. We did not detect any companion brighter than this limit, e.g., 4.5~mag at $1^{\prime\prime}$.

\subsubsection{Alopeke Speckle Imaging}
TOI-2257 was observed on 2021 February 02 UT using the ‘Alopeke speckle instrument on Gemini North \footnote {\url{https://www.gemini.edu/sciops/instruments/alopeke-zorro/}} \citep{Alopeke}. ‘Alopeke provides simultaneous speckle imaging in two bands (562 nm and 832 nm) with output data products including a reconstructed image with robust contrast limits on companion detections (e.g., \cite{howell2016}). 8 sets of 1000 X 0.06 sec exposures were collected and subjected to Fourier analysis in our standard reduction pipeline (see \cite{howell2011}). Figure \ref{fig:hires} shows our final contrast curves and the 832 nm reconstructed speckle image. We find that TOI-2257 is a single star with no companion brighter than about 4.5-5 magnitudes below that of the target star from the diffraction limit (20 mas) out to 1.2\arcsec. At the distance of TOI-2257 (d=57.8 pc) these angular limits correspond to spatial limits of 1\,au to 69\,au. 


\section{Stellar properties} \label{sec:stellar}
\begingroup
\begin{table}
\centering
\caption{Properties of the host star}

\begin{threeparttable}
\begin{tabular*}{\linewidth}{@{\extracolsep{\fill}}l c c@{}}
\toprule
Parameter & Value & Source \\
\midrule
& \textit{Designations} & \\
TIC & 198485881.01 & \\
2MASS & J12585767+7739416 & \\
Gaia DR2 & 1716345832872291968 & \\
UCAC 4 & 839-012174 & \\
\midrule
& \textit{Photometric magnitudes} & \\
TESS & 12.9672 $\pm $ 0.0074 & [1]\\
B & 16.648 $\pm$ 0.098 & [1]/[2] \\
V & 15.211 $\pm$ 0.034 & [1]/[2] \\
g' & 15.990 $\pm$ 0.242 & [2] \\
r' & 14.599 $\pm$ 0.057 & [2] \\
i' & 13.345 $\pm$ 0.177 & [2] \\
Gaia & 14.1615 $\pm$ 0.000437 & [3]\\
J & 11.47 $\pm$ 0.018 & [1]/[4] \\
H & 10.89 $\pm$ 0.015 & [1]/[4] \\
K & 10.673 $\pm$ 0.016 & [1]/[4] \\
WISE 3.4 $\mu m$ & 10.514 $\pm$ 0.023 & [5]\\
WISE 4.6 $\mu m$ & 10.355 $\pm$ 0.02 & [5]\\
WISE 12 $\mu m$ & 9.972 $\pm$ 0.039 & [5]\\
WISE 22 $\mu m$ & 8.601 $\pm$ 0.244 & [5]\\
\midrule
& \textit{Stellar properties} & \\
RA (J2000) & 12:58:57.51 & [1]\\
Dec (J2000) & +77:39:42.18 & [1]\\
pm (RA) $mas\, yr^{-1}$& $-36.035 \pm 0.021$ & [6] \\
pm (dec) $mas\, yr^{-1}$ & $31.408 \pm 0.018$ & [6] \\
Parallax $mas$ & $17.283 \pm 0.015$ & [6] \\
Distance $pc$ &  $57.7911^{+0.1053}_{-0.1049} $ & [7] \\
Spectral Type           & M3     & [8] \\ 
$T_{\text{eff}} /$ K   & 3430 \text{$\pm$} 130               & [8] \\ 
{[Fe/H]}                & -0.27 \text{$\pm$} 0.37   & [8] \\ 
$M_* / M_\odot$         & 0.33 \text{$\pm$} 0.02     & [8] \\ 
$R_* / R_\odot$         & 0.311 \text{$\pm$} 0.015     & [8] \\ 
$\log g$ / dex          & 4.971 \text{$\pm$} 0.050     & [8] \\ 
$\rho_*/\text{g cm}^{-3}$& 15.8 \text{$\pm$} 2.5      & [8]\\ 
$F_{\text{bol}}/ \text{erg s$^{-1}$cm$^{-2}$}$ & $(1.171 \pm 0.055) \times 10^{-10}$ & SED \\ 
\bottomrule
\end{tabular*}

\begin{tablenotes}
          \item 1: TIC \citep{Stassun2018}, 2: APASS-dr9 \citep{APASS9}, 3: Gaia-dr2 \citep{gaia18}, 4: Catalog of Cool Dwarf Targets \citep{muirhead2018}, 5: WISE \citep{WISE13}, 6: Gaia-dr3 \citep{GaiaEDR3}, 7: \citep{BailerJones2018}, 8: see Section \ref{sec:finalstellar}.
\end{tablenotes}

\end{threeparttable}
\label{tabl:starchar}
\end{table}
\endgroup

\subsection{Spectroscopic Analysis} 


Based on the Palomar/DBSP and NOT/ALFOSC spectra, we constrained the fundamental properties of TOI-2257 from the analysis of molecular band indices. 

We estimated the effective temperature $T_\mathrm{eff}$ employing the relationship between this parameter and the CaH2 index obtained for M dwarfs derived by \citet{Woolf2006}. We obtained $T_\mathrm{eff}$ = 3277 $\pm$ 250 K (CaH2 = 0.36) and $T_\mathrm{eff}$ = 3395 $\pm$ 310 K (CaH2 = 0.43) for the NOT/ALFOSC and Palomar/DBSP data, respectively. These values are in good agreement with those obtained from the optical
and infrared photometric calibration of \citet{Casagrande2008} which yields $T_\mathrm{eff}$ = 3279 $\pm$ 51 K for (V$-$Ks) and $T_\mathrm{eff}$ = 3266 $\pm$ 30 K for (V$-$J).

Based on the band strength indices CaH2, CaH3, and TiO5, we computed the coarse metallicity parameter $\zeta_{TiO/CaH}$ ($\zeta$ for short), as described by \citet{Lepine2007}. Then, we used the relation between this parameter and [Fe/H] derived by Mann et al. (2013) to obtain [Fe/H]= $-$0.29 $\pm$ 0.31 ($\zeta$ = 0.76 $\pm$ 0.3) and [Fe/H]= $-$0.23 $\pm$ 0.5 dex ($\zeta$ = 0.76 $\pm$ 0.3) for the NOT/ALFOSC and Palomar/DBSP data respectively. These values are in good agreement with the results obtained from photometric calibrations. We derived [Fe/H]= $-$0.35 $\pm$ 0.18 dex for the (V$-$Ks) calibration of \citet{Schlaufman2010}, [Fe/H]= $-$0.23 $\pm$ 0.22 dex for the (V$-$Ks) calibration of \citet{Mann2013}, and [Fe/H]= $-$0.23 $\pm$ 0.20 dex for the (Bp$-$Ks) calibration of \citet{Rains2021}.  

We derived several gravity-sensitive spectral indices to establish the main-sequence dwarf nature of TOI-2257. We computed the gravity-sensitive indices Na$_{8189}$ and TiO$_{7140}$ which several authors have shown that clearly separates low, intermediate, and high gravity for spectral types later than M2 \citep[e.g.,][]{Slesnick2006}. The measured spectral indices for the NOT/ALFOSC data are TiO$_{7140}$ = 1.40 and Na$_{8189}$ = 0.94, whilst for the Palomar/DBSP we obtained TiO$_{7140}$= 1.38 and Na$_{8189}$ = 0.91. These values place TOI-2257 in the high surface gravity region (dwarf stars) around the M3 spectral types in Fig. 11 of \citet{Slesnick2006}. As expected, a visual inspection of the spectra of TOI-2257 reveals a clear absorption by the NaI doublet at 8183 {\AA} and 8195 {\AA} which is not seen in giant stars. Furthermore, as in \citet{Alonso-Floriano2015}, we computed the index Ratio C, which also is highly sensitive to surface gravity. We derived Ratio C = 1.15 and Ratio C = 1.23 for the NOT/ALFOSC and Palomar/DBSP data, respectively. These values place TOI-2257 in the high surface gravity field-dwarf region in Fig. 6 of \citet{Alonso-Floriano2015}, which is well above from the limit of giant stars at Ratio C = 1.07. 

Finally, we assessed the spectral type of TOI-2257 using the PYHAMMER tool\footnote{\url{https://github.com/BU-hammerTeam/PyHammer}} \citep{Kesseli2017}, which estimates MK spectral types by comparing our data with reference spectra of M-type stars \citep{Covey2007}. The best fit is obtained for an M3V star with [Fe/H] $\sim$ 0 dex. Moreover, as in \citet{Slesnick2006}, we computed the temperature-sensitive index TiO$_{8465}$, in addition to the TiO$_{7140}$ index presented above, obtaining TiO$_{8465}$ = 1.16 and TiO$_{8465}$= 1.07 for the NOT/ALFOSC and Palomar/DBSP data, respectively. These values locate TOI-2257 around the M3 spectral type region in Fig. 9 of \citet{Slesnick2006}.



\subsection{SED fitting and evolutionary modeling}

\begin{figure}
    \centering
    \includegraphics[width=0.48\textwidth]{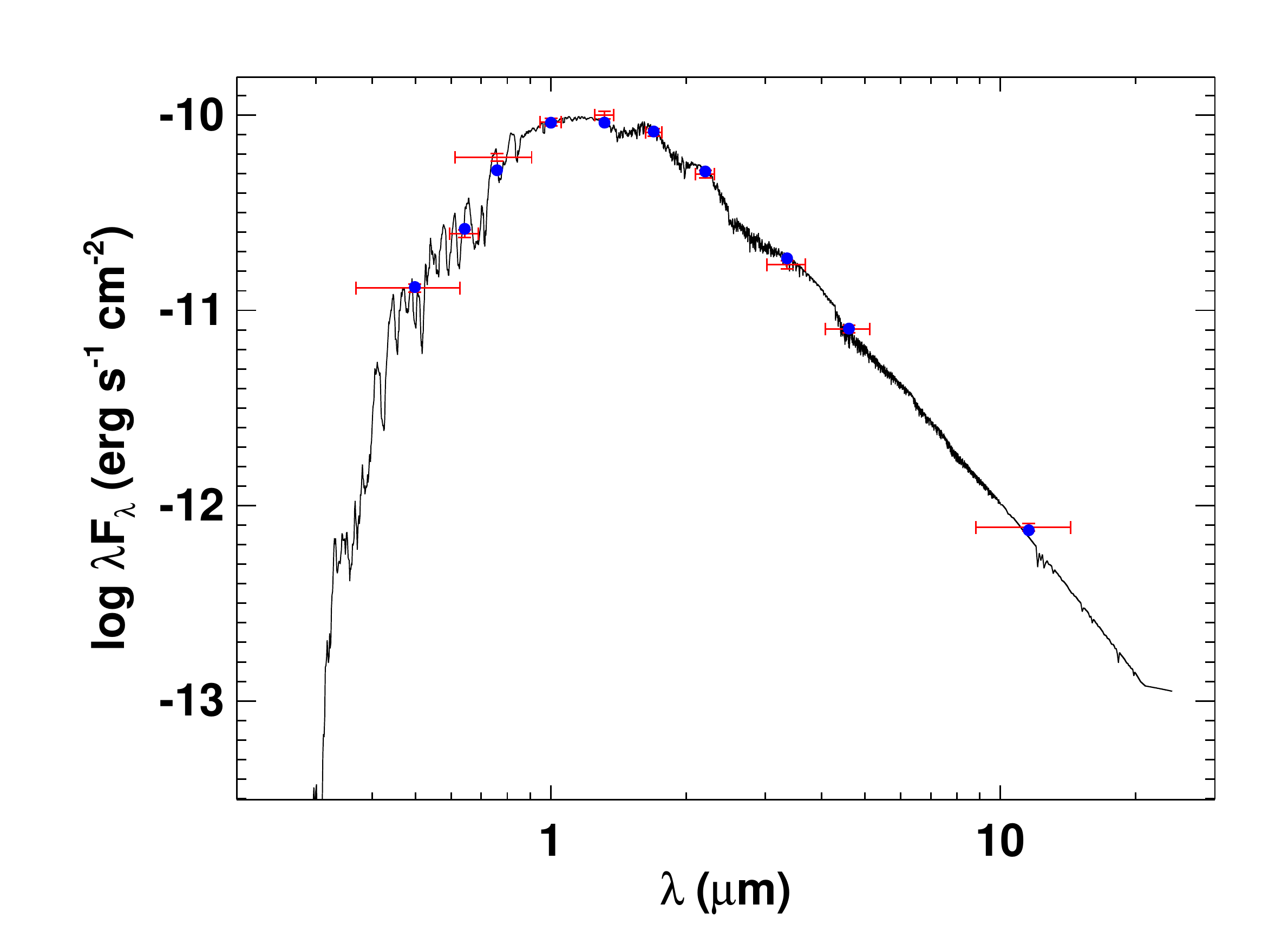}
    \caption{Spectral energy distribution of TOI-2257. The red symbols represent the observed photometric measurements;  the horizontal bars represent the effective width of the passband. The blue symbols are the model fluxes from the best-fit NextGen atmosphere model (black).}
    \label{fig:sed}
\end{figure}

As an independent determination of the basic stellar parameters, we performed an analysis of the broadband SED of the star together with the {\it Gaia\/} EDR3 parallax \citep[with no systematic correction][]{StassunTorres:2021} in order to determine an empirical measurement of the stellar radius, following the procedures described in \citet{Stassun:2016,Stassun:2017,Stassun:2018}. We pulled the $JHK_S$ magnitudes from {\it 2MASS}, the W1--W3 magnitudes from {\it WISE}, the $G G_{\rm BP} G_{\rm RP}$ magnitudes from {\it Gaia}, and the $gry$ magnitudes from Pan-STARRS. Together, the available photometry spans the full stellar SED over the wavelength range 0.4--10~$\mu$m (see Figure~\ref{fig:sed}).  

We performed a fit using NextGen stellar atmosphere models \citep{Hauschildt1999}, with the effective temperature ($T_{\rm eff}$), surface gravity ($\log g$), and metallicity ([Fe/H]) adopted from the spectroscopic analysis. The remaining free parameter is the extinction, $A_V$. The resulting fit (Figure~\ref{fig:sed}) has a reduced $\chi^2$ of 1.6, with best fit $A_V = 0.03 \pm 0.03$. Integrating the (unreddened) model SED gives the bolometric flux at Earth, $F_{\rm bol} = 1.168 \pm 0.041 \times 10^{-10}$ erg~s$^{-1}$~cm$^{-2}$. Taking the $F_{\rm bol}$ and $T_{\rm eff}$ together with the {\it Gaia\/} parallax gives the stellar radius  $R_* = 0.311 \pm 0.015$~R$_\odot$. Finally, the absolute $M_K$ magnitude together with the empirical relations of \citet{Mann:2019} imply a stellar mass of $M_* = 0.33 \pm 0.02$~M$_\odot$.
Together with the empirical radius above, this yields a mean stellar density of $\rho_* = 15.7 \pm 2.5$~g~cm$^{-3}$ and a surface gravity of $\log g = 4.971 \pm 0.050$.

We also estimated the stellar properties using a complementary, isochrone-dependent approach.
For this analysis, we used the \texttt{isochrones} software package \citep{isochrones}, which can be used to fit data inputs to the MESA Isochrones and Stellar Tracks database \citep{MIST0, MIST1} via the nested sampling algorithm MULTINEST \citep{MULTINEST} as implemented in the \texttt{PyMultiNest} package \citep{PyMultiNest}.
We used as inputs the stellar metallicity from \autoref{tabl:starchar}, the $BV$ magnitudes tabulated in the revised TESS Input Catalog \citep[TIC,][]{TICv8}, the $JHK_s$ magnitudes from 2MASS \citep{2MASS, 2MASS_PSC}, and the W1--W3 magnitudes from WISE \citep{WISE} as well as the \textit{Gaia} EDR3 parallax \citep{Gaia2016_mission, GaiaEDR3}.
This analysis gives 
a stellar effective temperature of $T_\mathrm{eff} = 3441 \pm 19$\,K,
surface gravity of $\log g = 4.926 \pm 0.054$,
mass of $M_* = 0.315 \pm 0.028\,M_\sun$,
radius of $R_* = 0.3190 \pm 0.0024$, and
density of $\rho_* = 13.7 \pm 1.3$\,g\,cm$^{-3}$ in line with the results of the SED fitting.
The results also favor a stellar age of $\ge 8.0$\,Gyr at 3$\sigma$ and minimal line-of-sight extinction ($A_V < 0.2$ at $3\sigma$).
While we note the consistency of the isochrone-derived parameter values with the others presented here, we do not include them in the calculation of the final, adopted stellar parameters as the uncertainties are likely underestimated.

\subsection{Adopted Stellar Parameters} \label{sec:finalstellar}

We compiled our final set of adopted stellar parameters from the analyses of the NOT/ALFOSC and Palomar/DBSP spectra and the SED.
In particular, the analyses detailed previously yielded three estimates of $T_\mathrm{eff}$ (from NOT/ALFOSC, Palomar/DBSP, and the SED analyses) and two estimates of [Fe/H] (from the NOT/ALFOSC and Palomar/DBSP analyses).
For each parameter, we calculated the weighted mean of the estimates using a Monte Carlo approach.
We drew a total of $10^6$ samples from normal distributions with the means and standard deviations of the estimates, using inverse-variance weighting to determine the number of subsamples to draw from each distribution.
We then calculated the mean and standard deviation of the resulting sample.
We adopted these values as our final estimates of $T_\mathrm{eff}$ and [Fe/H], which are presented in \autoref{tabl:starchar}.


\section{Transit analysis} \label{sec:fit}
In this section we outline the fit of the data to a transit model as well as an analysis of possible transit timing variations (TTVs).

\subsection{Transit model fit}
All the lightcurves described in Section \ref{sec:obs} were used simultaneously to fit the transit using a Markov Chain Monte Carlo (MCMC) approach. We employed the \texttt{PyTransit} \citep{Parviainen2015_pytransit} implementation of the \cite{Mandel2002} quadratic limb darkening transit model, with sampling done using \texttt{emcee} \citep{ForemanMackey2013}, a Python implementation of the Affine Invariant MCMC Ensemble sampler \citep{Goodman2010}. 
The parameters used for the transit model are the planet-to-star radius ratio $R_p/R_*$, impact parameter $b$, zero epoch $T_0$, period $P$, and stellar mass and radius $M_*$ and $R_*$. The eccentricity and argument of periastron were fit using the parameterization of $\sqrt{e}\cos{\omega}$ and $\sqrt{e}\sin{\omega}$. In addition, two quadratic limb darkening parameters were fit for each wavelength. 
The ground-based data was simultaneously detrended using an out-of-transit baseline as well as quadratic terms for the airmass and FWHM trends. Priors used for the analysis are shown in Table \ref{table:planet_params}.

\begin{figure*}
    \centering
    \includegraphics[width=0.75\textwidth]{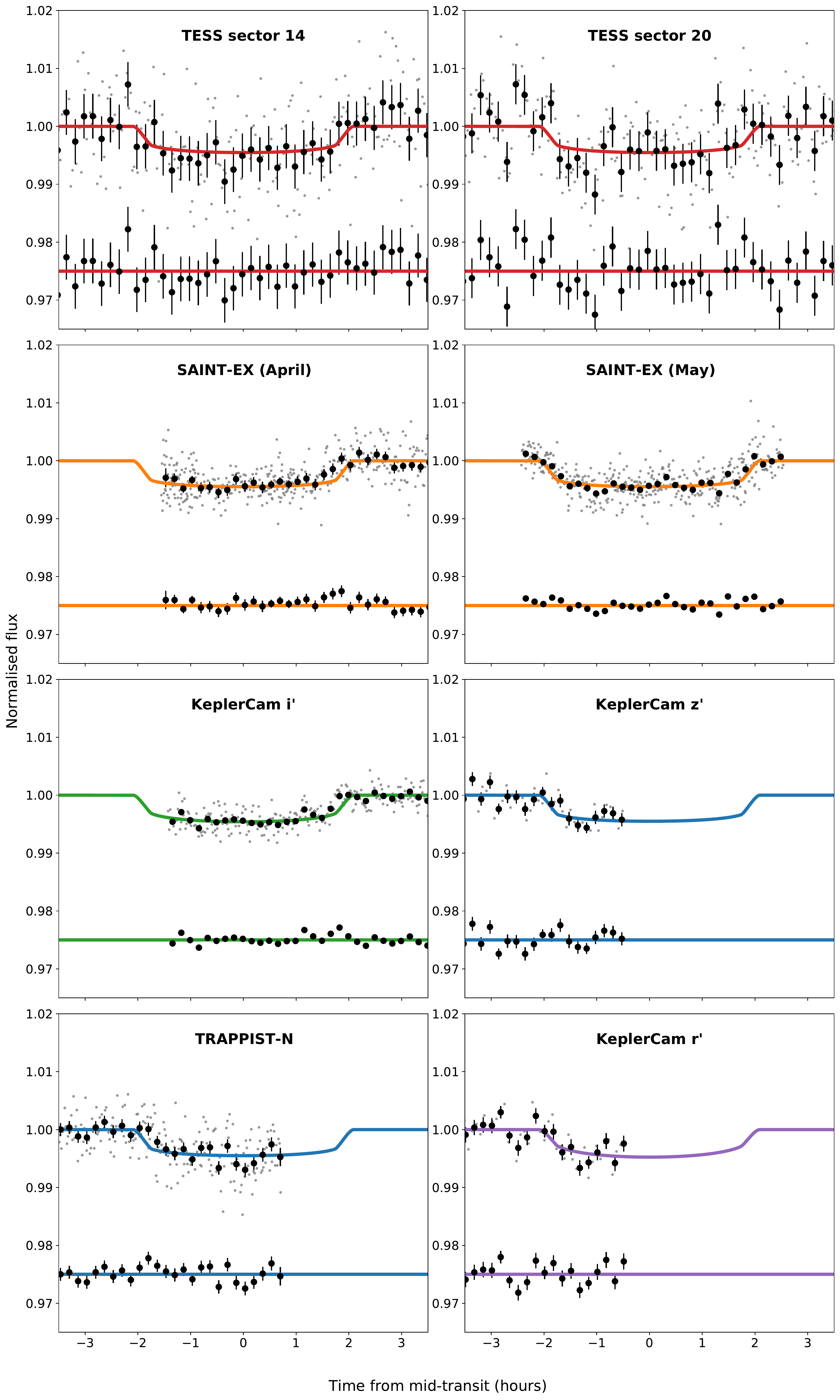}
    \caption{Photometric observations of TOI-2257 with the best-fit transit parameters over-plotted. Thick black points show 10-minute bins, with all data points shown in gray. Residuals from the model fit are shown with an arbitrary offset below the transit. Note only 10-minute binned residuals are shown for clarity. The color of the line reflects the filter used for the observations (red for TESS, orange indicates I+z, green is \textit{i}', blue for \textit{z}', and purple for \textit{r}').} 
    \label{fig:all_transit_fit}
\end{figure*}

The final transit model is shown in Figure \ref{fig:all_transit_fit} over the detrended lightcurves. The posterior distributions for the transit parameters can be found in Appendix \ref{fig:mcmc}, and the final fit and derived parameters for the planetary system can be found in Table \ref{table:planet_params}. While we currently have no mass estimations from radial velocity measurements, we use the relationship in \cite{Chen2017} to estimate a range of likely values. The radial velocity semi-amplitude thus can be estimated, leading to an expected signal around 3.5\,m\,s$^{-1}$. 

We also carried out an independent analysis using the \texttt{juliet} package \citep{2019MNRAS.490.2262E}, which is built over \texttt{batman} \citep{kreidberg_bbatman_2015} for the modeling of transits and the \texttt{dynesty} \citep{2020MNRAS.493.3132S} dynamic nested sampling algorithm for estimating Bayesian posteriors and evidences. The fitted transit parameters were: $R_p/R_*$, $b$, $T_0$, $P$, the stellar density $\rho_*$, $\sqrt{e}\cos{\omega}$, and $\sqrt{e}\sin{\omega}$. For each passband, we also fitted two quadratic limb-darkening coefficients, which were parameterized using the ($q_1$, $q_2$) triangular sampling scheme of \cite{exoplanet:kipping13}. All these parameters were sampled from wide uniform priors, except the stellar density, for which we used a normal prior based on the value and uncertainty reported in Table \ref{tabl:starchar}.

We first performed individual analyses of each of our lightcurves, in order to select for each of them the best correlated noise model, based on Bayesian evidence. We explored a large range of models, consisting of first- to fourth-order polynomials with respect to, e.g., time, airmass, PSF
FWHM, background, stellar position on the detector, or any combination of these parameters. First- or second-order polynomials of airmass and/or FWHM were typically favored. For each lightcurve, we also fitted a jitter term, which was added quadratically to the error bars of the data points, to account for any underestimation of the uncertainties or any excess noise not captured by our modeling.

We then conducted two global analyses: one assuming a circular orbit ($e$ set to zero) and one allowing the orbit of the planet to be eccentric. Following \cite{2019MNRAS.490.2262E}, we found the eccentric model ($e = 0.41_{-0.12}^{+0.21}$) to be strongly favored over the circular one, with a difference in Bayesian log evidence ($\Delta \ln Z$) of 5.2, i.e. posterior odds of $\approx$180 : 1 (assuming equiprobable models) lending substantial support to our eccentric interpretation \citep{Kass1995}. All transit parameters returned by the eccentric fit are consistent within the uncertainties with the ones reported in Table \ref{table:planet_params}.

\begin{table*}
\centering
\caption{Fitted and derived parameters for the TOI-2257\,b system.}
\begin{threeparttable}
\begin{tabular}{l c c r}
\toprule
Parameter & Unit & Value & Prior  \\
\midrule
\multicolumn{4}{c}{\it{Fitted Parameters}} \\
\hline
Orbital period  $P$ & days &  $35.189346(90)$ & $\mathcal{N}\,(35.189295, 1e-4)$ \\
Mid-transit time $T_0$ & BJD-2450000 &  $9007.97949^{+0.00108}_{-0.00105}$ & $\mathcal{N}\,(9007.978906,  0.1)$ \\
$(R_p / R_*)$ & &  $0.06423^{+0.00142}_{-0.00133}$ & $\mathcal{U}\,(0.001, 0.4)$ \\
Impact parameter $b$ & &  $0.374^{+0.098}_{-0.137}$ & $\mathcal{U}\, (0,1)$ \\
$\sqrt{e}sin\omega$ & &  $-0.615^{+0.083}_{-0.073}$ & $\mathcal{U}\, (-1,1)$ \\
$\sqrt{e}cos\omega$ & &  $-0.126^{+0.484}_{-0.387}$ & $\mathcal{U}\, (-1,1)$ \\
Stellar Mass $M_*$ & $M_\odot$ &  $0.328^{+0.021}_{-0.019}$ & $\mathcal{N}\,(0.33,0.02)$\\
Stellar Radius $R_*$ & $R_\odot$ &  $0.313 \pm 0.015$ & $\mathcal{N}\,(0.311, 0.015)$\\
\hline
\multicolumn{4}{c}{\it{Physical and Orbital Parameters}} \\
\hline
Planet Radius $R_p$ & $R_\oplus$ & $2.194^{+0.113}_{-0.111}$ & \\
Orbital Eccentricity $e$ & &  $0.496^{+0.216}_{-0.133}$ & \\
Argument of Periastron $\omega$ & $^{\circ}$ &  $-101.674^{+42.453}_{-26.881}$ & \\
Semi-major axis $a$ & au &  $0.145 \pm 0.003$ & \\
Inclination $i$ & $^{\circ}$ &  $89.786^{+0.078}_{-0.062}$ & \\
Equilibrium Temperature $T_{eq}$\tnote{*} & K & $256^{+61}_{-17}$ & \\
Depth $\delta$ & &  $0.00413^{+0.00018}_{-0.00017}$ & \\
Transit duration & hrs &  $3.846^{+0.057}_{-0.051}$ & \\
\hline
\multicolumn{4}{c}{\it{Predicted Parameters}} \\
\hline
Planet Mass $M_p$ & $M_\oplus$ &  $5.712^{+4.288}_{-2.311}$ & \\
RV Semi-amplitude $K$ & $m\, s^{-1}$ & $3.521^{+2.901}_{-1.507}$ & \\
TSM & & $32.708^{+24.08}_{-14.362}$ & \\
\hline
\multicolumn{4}{c}{\it{Limb Darkening}} \\
\hline
q1 ($TESS$) & &  $0.328 \pm 0.008$ & $\mathcal{N}\,(0.3281,0.0079)$ \\
q2 ($TESS$) & &  $0.203 \pm 0.022$ & $\mathcal{N}\,(0.2020,0.0229)$ \\
q1 ($I+z$) & &  $0.289 \pm 0.008$ & $\mathcal{N}\,(0.2903,0.0076)$ \\
q2 ($I+z$) & &  $0.193^{+0.027}_{-0.026}$ & $\mathcal{N}\,(0.1992,0.0260)$ \\
q1 ($i'$) & &  $0.385^{+0.01}_{-0.009}$ & $\mathcal{N}\,(0.3849,0.0098)$ \\
q2 ($i'$) & &  $0.213 \pm 0.03$ & $\mathcal{N}\,(0.2070,0.0301)$ \\
q1 ($z'$) & &  $0.300^{+0.008}_{-0.009}$ & $\mathcal{N}\,(0.2994,0.0088)$ \\
q2 ($z'$) & &  $0.206^{+0.034}_{-0.031}$ & $\mathcal{N}\,(0.2050,0.0327)$ \\
q1 ($r'$) & &  $0.568^{+0.012}_{-0.011}$ & $\mathcal{N}\,(0.5686,0.0116)$ \\
q2 ($r'$) & &  $0.164 \pm 0.034$ & $\mathcal{N}\,(0.1663,0.0326)$ \\
\midrule
\end{tabular}
\begin{tablenotes}\footnotesize
\item[*] Time averaged equilibrium temperature taking into account the eccentric orbit using the formulation of \citep{Mendez2017} and assuming zero albedo and full day-night heat redistribution. 
\end{tablenotes}
\end{threeparttable}
\label{table:planet_params}
\end{table*}

\subsection{Transiting Timing Variations}

TTVs are of interest in this context because they can point to the existence of other planets in the system and potentially be used to constrain the mass of the transiting planet.
We investigated the evidence for TTVs of TOI-2257\,b in a separate analysis of the transit lightcurves, generating the TTV lightcurve model with the \texttt{exoplanet} software package \citep{exoplanet:joss} and performing the posterior inference with the \texttt{PyMC3} software package \citep{exoplanet:pymc3}.
Our approach followed similarly that of other recent TTV analyses of TESS-discovered exoplanets \citep{Badenas-Agusti2020, Daylan2021}.

As data inputs, we used the TESS Sector~14 and 20 2-minute-cadence lightcurves, the 20\,Apr\,2021 partial-transit lightcurves from SAINT-EX, KeplerCam, and TRAPPIST-N, and the 25\,May\,2021 full-transit lightcurve from SAINT-EX, the most constraining data on this particular transit.
We placed normal priors on the stellar mass and radius using the values presented in \autoref{tabl:starchar} and on the planet-to-star radius ratio and period using the values presented in \autoref{table:planet_params}.
We placed uninformative priors on the impact parameter and the quadratic limb-darkening coefficients using the physical distributions built into \texttt{exoplanet}, the latter of which relies on the triangular sampling method of \citet{exoplanet:kipping13}.
We placed a normal prior on the period of the planet using the results of the linear-ephemeris analysis and uniform priors on the times of the four observed transits, centered on the expected transit times from the linear ephemeris and with a width of 1\,hr.
Finally, we placed a normal prior on the out-of-transit flux for all datasets, centered on the median flux value and using a standard deviation that is robustly estimated using the median absolute deviation of the fluxes.

This model has some notable simplifications.
We assume that the noise is normally distributed and do not attempt to model any residual systematics in the normalized lightcurves.
Since all the observations are in the red optical, generally, we also use a single set of limb-darkening coefficients and a single transit depth to model the lightcurve despite the somewhat different bandpasses of the observations.
Finally, we use a single out-of-transit flux value for the combined dataset, assuming that any errors in the normalization of the individual datasets are negligible.
These simplifications notably reduce the complexity of the model and improve the computational efficiency of the sampling, and we assume that they do not significantly impact the inference of the transit times.

We find that the four transits are consistent with a linear ephemeris at the $1\sigma$ level with inferred TTVs of $1.5\pm4.7$, $-2.0\pm6.4$, $0.3\pm1.1$, and $0.3\pm1.2$ minutes. The smaller uncertainties of the last two transits illustrate the improved timing estimates afforded by our ground-based follow-up observations.
We conclude that the existing observations do not show any evidence of TTVs. 
TESS observations of TOI-2257 in the near future (in Sectors 40, 41, 47, 48, and 53) can be monitored to check if any potential variations emerge.


\section{Planet validation} \label{sec:validation}

\subsection{TESS Data Validation Report}
The natural first step for false-positive vetting was to closely analyze the TESS data validation report \citep{Twicken2018, Li2019} based on the 176-day solution. All tests, including those for odd-even transit depths, centroid shifts, and `ghosts', produced favorable results. The encouraging results led to further follow-up in the system.

\subsection{Follow-up photometry}\label{}
The TESS DV report indicates that there is one contaminating source within the TESS apertures used in both sectors 14 and 20 at a distance of 16.47". We identify this as the faint source with Gaia ID 1716345832871506560 (G-mag 20.7). The pixel scale from the ground-based follow-up photometry extracted lightcurves using an aperture of only a few arcseconds, and therefore we were able to resolve the target in isolation. No transit event was seen on this faint star or any other nearby star, while a dimming event was clearly observed on the target. 

\subsection{Archival imaging} \label{sec:archival}

\begin{figure*}
    \centering
    \includegraphics[width=\textwidth]{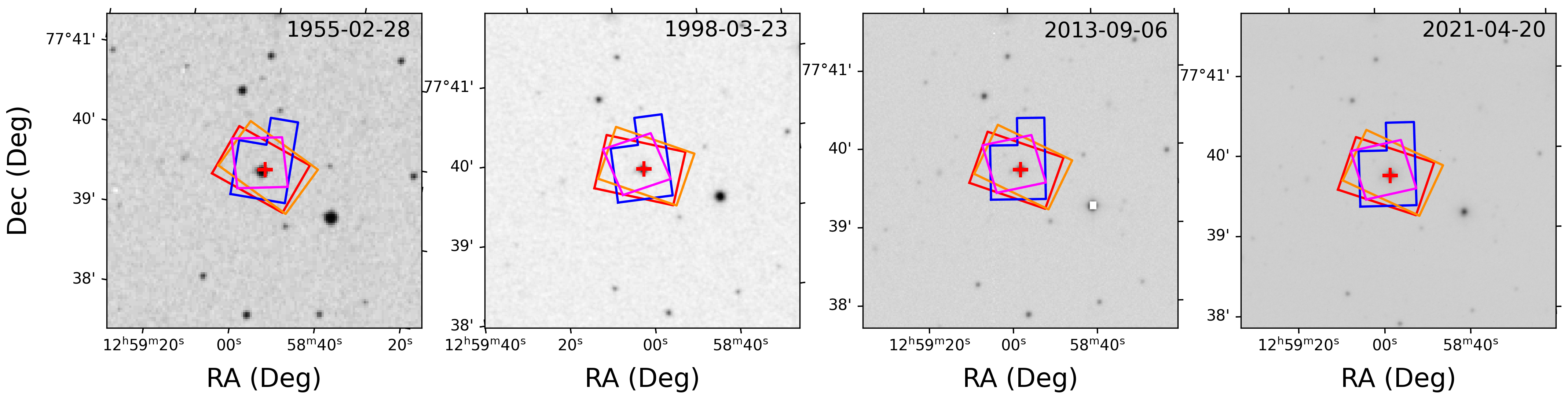}
    \caption{Archival images around TOI-2257 with TESS's apertures used in Sector 14 (red), 20 (blue), 21 (yellow) and 26 (magenta) superimposed to assess for current, unresolved blending. From the left to the right: (1) 1955-02-28 DSS1-red filter, (2) 1998-03-23 DSS2-red filter, (3) 2013-09-06 PTF-red filter, and (4) 2021-04-20 TN-red filter. 
    Its current location (red cross) is marked in all images.}
    \label{fig:archive}
\end{figure*}

We used archival images to investigate the contamination of background stars \citep[e.g.,][]{quinn2019}. We aimed to rule out the presence of EBs at the present-day target location, which might introduce transit-like signals in the data. Unfortunately, TOI-2257 has a moderately low proper motion of $\sim$0.047 arcsec~yr$^{-1}$. The oldest archival image that we found was taken in 1955, 64 and 66 years before the first and the last TESS observations in Sector 14 and 26, respectively. This oldest archival image has a pixel scale and point-spread-function (PSF) of 1.69~arcsec/pixel and 8.05~arcsec, respectively. Since that time, the star only moved $\sim$3.1 arcsec (see Fig. ~\ref{fig:archive}). While it seems that there is not any background star that might be producing the transit-like signals detected in our data, we can not confidently rule out that possibility. We would need to wait for $\sim$150~years to reach a separation of 10~arcsec since 1955, a distance larger than the 1955's PSF that would allow us to rule out confidently the contamination caused by a background EB.

\subsection{False-positive likelihood} \label{sec:fpp}
To assess the possibility that the observed transit was not due to a planet orbiting the target star, but rather from an astrophysical false positive scenario, we used the software package \texttt{triceratops} \citep{Giacalone2021}. This package was developed as a tool to assist in the vetting and validation of TESS candidates (See, e.g., \cite{Cloutier2020,demory:2020, Hedges2021, wells2021}). Using a Bayesian framework, \texttt{triceratops} calculates the probability that the signal is caused by a variety of true positive or false positive scenarios. The calculations incorporate prior knowledge about the target star, exoplanet occurrence rates, and stellar multiplicity. One metric that is returned is the False Positive Probability (FPP), which is the sum of probabilities for all false positive scenarios. Using the TESS 2-minute data from sector 14 and sector 20 gives an FPP of $0.078\pm0.009$. However, \texttt{triceratops} also is able to incorporate additional information from contrast curves to further constrain the false positive scenarios. By incorporating this information, the FPP is reduced to $0.0255\pm0.0012$. However, the higher precision of the SAINT-EX lightcurve is more constraining than the TESS data. We therefore modify the input to \texttt{triceratops} to use the detrended SAINT-EX transit observation from May 25 along with the contrast curve, reducing the FPP to a negligible value ($\sim 8.7 \times 10^{-9}$). We therefore are able to statistically validate this object as a planet. 

\subsection{Unresolved stellar companions}

 Based on the high-resolution imaging (see Section \ref{sec:HiRes}), we ruled out the potential for a companion within 4.5-5.5 mag of the target star outside ~6 au at 832 nm, which corresponds to a magnitude of ~18.6 in the r' band. Theory and observations established that the mass cut-off for what constitutes a star is different for objects of different metallicity. For objects with a Solar-like metallicity, anything with less than 0.075 M$_{\odot}$ will be a brown dwarf \citep{boss2001}, while for objects with lower metallicity, the mass limit will be about 0.083 M$_{\odot}$ \citep{richer2006}. Using the models by \cite{Baraffe2015} and assuming a stellar age of 5\, Gyr, we found an upper limit for the companion's mass of 0.075\,M$_\odot$. Hence, this result allows us to confidently rule out the presence of an unresolved stellar companion. Nevertheless, sub-stellar objects with masses ranging from 0.01--0.075\,M$_{\odot}$ would still be possible.


\section{Planet searches and detection limits}\label{sec:search}
\subsection{TESS photometry}

To search for additional planets, we used our custom pipeline {\fontfamily{pcr}\selectfont  SHERLOCK}\footnote{{The \fontfamily{pcr}\selectfont  SHERLOCK} (\textbf{S}earching for \textbf{H}ints of \textbf{E}xoplanets f\textbf{R}om \textbf{L}ightcurves \textbf{O}f spa\textbf{C}e-based see\textbf{K}ers) code is fully available on GitHub: \url{https://github.com/franpoz/SHERLOCK}} \citep{pozuelos:2020,demory:2020}. {\fontfamily{pcr}\selectfont  SHERLOCK} is a user-friendly open-source package that has five different modules that allow the user to: (1) search for planetary candidates; (2) perform vetting of the most promising signals; (3) compute a statistical validation; (4) model the signals to refine the ephemerides; and (5) compute the observational windows from ground-based observatories. {\fontfamily{pcr}\selectfont  SHERLOCK} has direct access to short- and long-cadence data observed by Kepler/K2 and TESS. Hence, {\fontfamily{pcr}\selectfont  SHERLOCK} is a fully operational, powerful tool that allows users to perform the planet search fast and robustly. 

{\fontfamily{pcr}\selectfont  SHERLOCK} applies a multi-detrend approach to the nominal lightcurve employing the \texttt{w{\={o}}tan} package \citep{wotan:2019}, that is, the nominal lightcurve is detrended several times using a bi-weight filter by varying the window size. This strategy allows the user to maximize the signal detection efficiency (SDE) and the signal-to-noise ratio (SNR) of the transit search, which is performed over the nominal lightcurve, jointly with the new detrended lightcurves, through the \texttt{transit least squares} (TLS) package \citep{hippke_TLS_2019}. TLS is optimized to detect shallow periodic transits using an analytical transit model based on the stellar parameters. The transit search is performed in a loop; once a signal is found, it is recorded and masked, then the search keeps running until no more signals with SNR$\geqslant$5 are found in the data set. 
To start the search for extra planets, we masked the transits corresponding to the candidate TOI-2257\,b, with an orbital period of 35.19~d, T$_{0}$=1691.28~d and T$_{14}$=228~min. Then, we performed three transit searches; firstly, by considering all the sectors available simultaneously, that is, combining sectors 14, 20, 21 and 22. We focus our search for orbital periods ranging from 0.5 to 30\,d, where at least two transits were required to identify a potential signal. Secondly, we focused on the longer orbital periods, ranging from 40 to 80\,d. In this case, we allowed recovering single transits. Finally, we explored all the sectors independently, focusing on orbital periods ranging from 0.5 to 15\,d. This strategy allowed us to avoid that sectors with different photometric precision affect the global search.

After the scrutiny of the data, we found no clear evidence of additional planetary transits. All the signals found by {\fontfamily{pcr}\selectfont  SHERLOCK} were attributable to systematics, noise, or variability. Following \cite{wells2021}, the lack of detecting extra signals suggests that: (1) no other planets exist in the system; (2) if they do exist, they do not transit; or (3) they exist and transit, but the photometric precision of the data set is not good enough to detect them, or they have longer periods than the ones explored in this study. 
If scenario (2) or (3) is true, extra planets might be detected by radial velocity follow-up, as discussed in Section~\ref{sec:discussion}. 

To evaluate scenario (3), we studied the detection limits of the current data set by performing injection-and-recovery experiments over the PDC-SAP lightcurves, combining the four sectors available. To this end, we  used the {\fontfamily{pcr}\selectfont  MATRIX ToolKit} \footnote{{The \fontfamily{pcr}\selectfont  MATRIX ToolKit} 
(\textbf{M}ulti-ph\textbf{A}se \textbf{T}ransits \textbf{R}ecovery from \textbf{I}njected e\textbf{X}oplanets \textbf{T}ool\textbf{K}it) code is available on GitHub: \url{https://github.com/martindevora/tkmatrix}}.

{\fontfamily{pcr}\selectfont  MATRIX ToolKit} allows the user to define the ranges in the $R_{\mathrm{planet}}$--$P_{\mathrm{planet}}$ parameter space to examine. Each combination of $R_{\mathrm{planet}}$--$P_{\mathrm{planet}}$ is explored using a number of different phases, that is, different values of T$_{0}$. For simplicity, it is assumed the impact parameters and eccentricities of the injected planets are zero. To perform the injection-and-recovery experiments, {\fontfamily{pcr}\selectfont  MATRIX ToolKit} injected the synthetic planets, detrended the lightcurves using a biweight filter with a window-size of 0.95~d, which was found to be the optimal value during the {\fontfamily{pcr}\selectfont SHERLOCK}'s runs, and masked the transits corresponding to the candidate TOI-2257\,b. We considered a synthetic planet to be recovered when its epoch was detected with 1~h accuracy, and the recovered period was within 5\,\% of the injected period. It is worth noting that since we injected the synthetic planets in the PDC-SAP lightcurve, the signals were not affected by the PDC-SAP systematic corrections; therefore, the detection limits should be considered as the most optimistic scenario \citep[see e.g.][]{pozuelos:2020,eisner:2019}.

\begin{figure}
    \centering
    \includegraphics[width=0.48\textwidth]{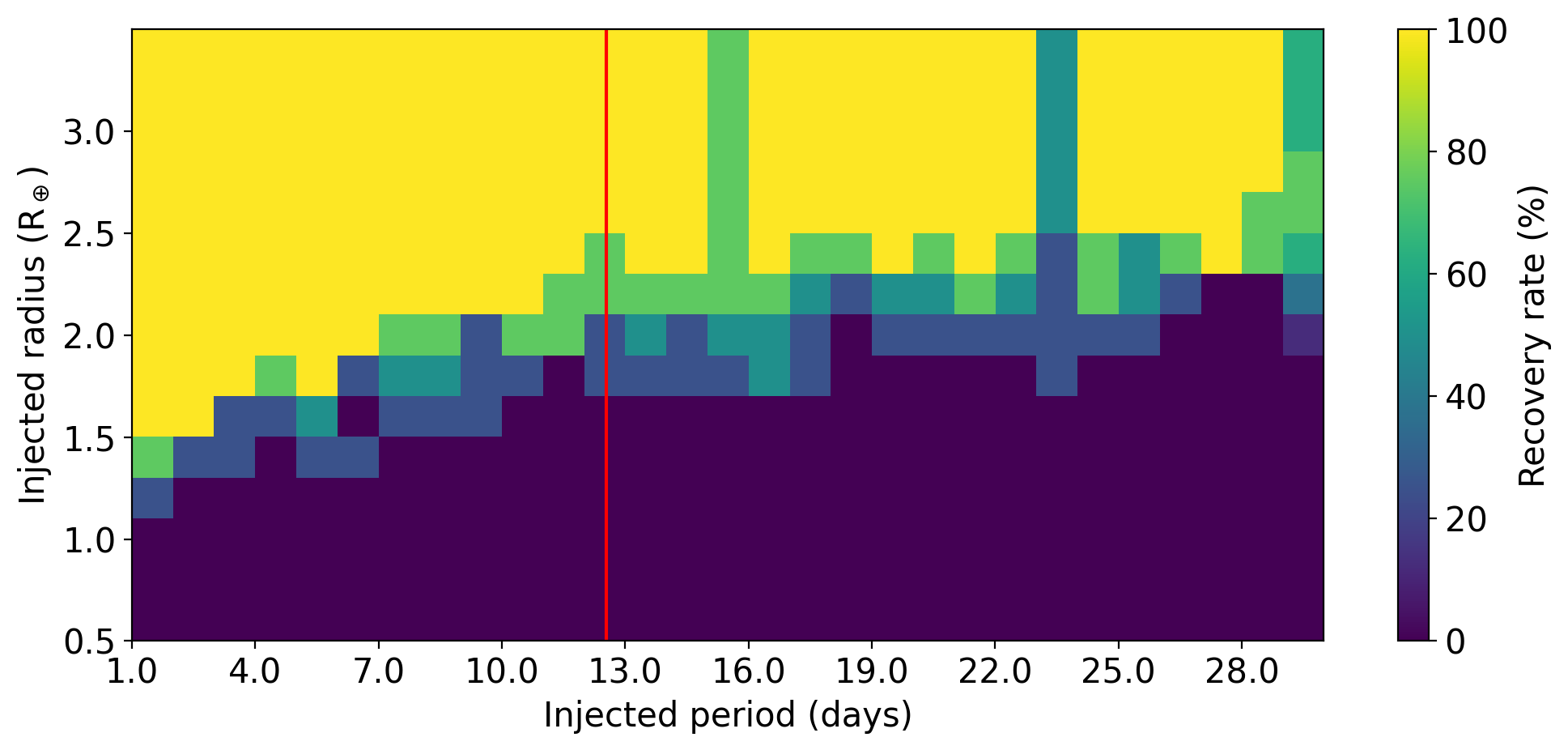}
    \caption{Injection-and-recovery tests performed on TESS data to check the detectability of extra planets in the system. We explored a total of 1920 different scenarios. Each pixel evaluated four scenarios, that is, four lightcurves with injected planets having different $P_{\mathrm{planet}}$, $R_{\mathrm{planet}}$ and T$_{0}$. Larger recovery rates are presented in yellow and green colors, while lower recovery rates are shown in blue and darker hues. We can rule out the presence of planets with sizes $>$2.5\,R$_{\oplus}$. Planets with sizes $<$1.5\,R$_{\oplus}$ would remain undetected. Planets with sizes between 1.5 and 2.5 $R_{\oplus}$ have recovery rates ranging from 30 to 70\,$\%$. The red vertical line marks the maximum orbital period (12.54~d) allowed for the hypothetical inner planet to avoid crossing orbits between the planets. See text for more details.}
    \label{fig:recovery}
\end{figure}

In particular, we explored the $R_{\mathrm{planet}}$--$P_{\mathrm{planet}}$ parameter space in the ranges of 0.5 to 3.5\,R$_{\oplus}$ with steps of 0.2\,R$_{\oplus}$, and 1.0--30.0\,d with steps of 1.0\,d. For each pair $R_{\mathrm{planet}}$--$P_{\mathrm{planet}}$ we used four different phases. Hence, we analyzed a total of 1920 scenarios. The results are shown in Fig.~\ref{fig:recovery}, which allowed us to rule out planets with sizes $>$2.5\,R$_{\oplus}$, with recovery rates larger than 80\,$\%$ almost for the full range of periods explored. On the other hand, planets with sizes $<$1.5\,R$_{\oplus}$ would remain undetected. In addition, we found that planet sizes $1.5<R_{\oplus}<2.5$ might be challenging to detect, with recovery rates ranging from 30 to 70\,$\%$.

It is important to notice that with the current data set, our models for TOI-2257\,b presented in Section~\ref{sec:fit} favored an eccentric orbit. Then, there are forbidden periods for the hypothetical inner planet to avoid crossing orbits between the planets. In particular, 
safe orbits have a semi-major axis lower than the periastron distance for TOI-2257\,b ($q$=0.073~au). That is, the allowed orbital periods would be $\lesssim$12.54~d. This limit in the orbital period for the hypothetical inner planet is displayed in Fig.~\ref{fig:recovery} with a vertical red line.

\subsection{SAINT-EX + Artemis photometry}\label{sec:speculoos_photometry}

\begin{figure}
    \centering
    \includegraphics[width=1\columnwidth]{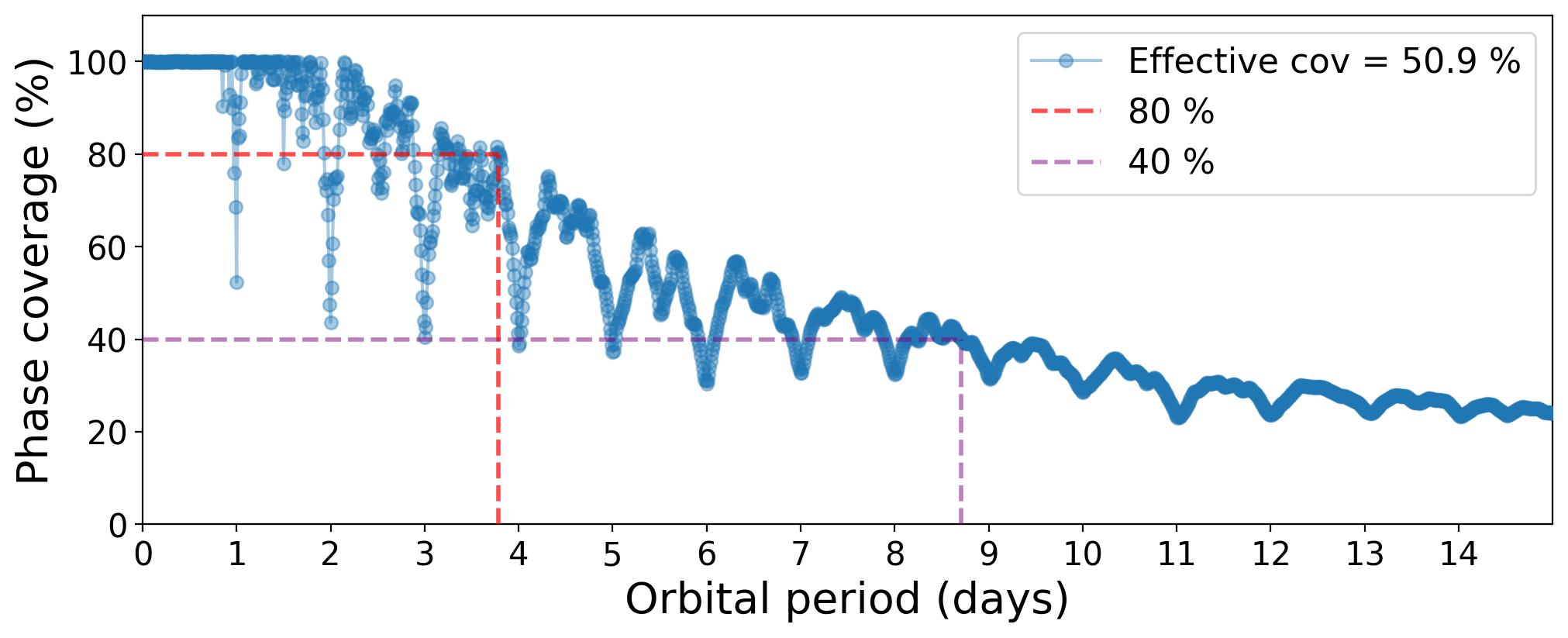}
    \includegraphics[width=0.48\columnwidth]{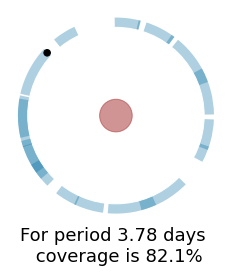}
    \includegraphics[width=0.48\columnwidth]{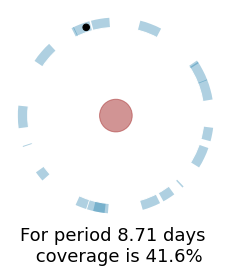}
    \caption{{\it Top:}Evolution of the phase coverage of a hypothetical planet orbiting TOI-2257 as a function of the period, derived from SAINT-EX and Artemis observations (blue dots). The effective phase coverage is the integral of the phase coverage over the 0.1 to 15 day period range and is equal to 59\%. The dotted red/purple lines indicate
    the period above which the phase coverage is always inferior to 80\% (40\%), which corresponds to the periods 3.78 days and 8.71 days respectively. We note that periods equal to an integer number of days are significantly less covered due to day-night cycles in ground-based observations.
    {\it Bottom:} Graphical visualisation of the coverage of TOI-2257 with SNO and SAINT-EX for an hypothetical planet with orbital period of 3.78 and 8.71 days respectively. Each blue circular arc represents one night of observation; its size is proportional to the number of hours observed each night and the full circle depicts a duration of 3.78 (8.71) days.}
    \label{fig:phase}
\end{figure}

As shown previously using TESS photometry, it would be challenging to find small planets with sizes $\leq1.5$~R$_{\oplus}$. Accordingly, we monitored TOI-2257 using the SPECULOOS network located in the Northern Hemisphere. This high-precision photometry allows us to detect single transits of Earth-size planets orbiting M-dwarfs \citep{delrez_speculoos_2018,murray_photometry_2020,demory:2020,Niraula2020}. Outwith the two transits shown in Section~\ref{saintex} we followed up the star for 93.5~h. These time-series observations were analyzed by our automatic pipelines as described in \cite{murray_photometry_2020} and \cite{demory:2020} in the search for transit events. 
We did not find any hint of transit-like features during this time. Hence, this negative result allowed us to rule out the presence of Earth-size planets for a range of orbital periods by computing the phase coverage; that is, we computed the percentage of phase covered for each orbital period from 0.1~d to 15~d, in intervals of 0.01~d, for a total of 1500 periods. We found that because of day-night cycles only very short periods are fully covered in phase by our observations, as shown in Fig.~\ref{fig:phase}. Indeed, only for orbital periods $\leq$3.78~d, more than ~80$\%$ of the orbit was explored, meaning that periods equal to or shorter than this would most likely be detected if the planet exists and transits. For larger orbital periods, the phase coverage decreases rapidly to a minimum of 20$\%$ for the highest orbital period explored of 15~d. 

\section{Discussion } \label{sec:discussion}

\subsection{Composition}
The radius of TOI-2257\,b is $2.194^{+0.113}_{-0.111}\,R_{\oplus}$. Without RV measurements, we use the relation of \cite{Chen2017} to estimate the likely mass, providing a wide range of predicted values centered around 5.7$M_\oplus$. Using mass-radius curves from \cite{Zeng2016}, it is expected that the planet's density is consistent with a composition of an ice/gas giant rather than a denser rocky body. 


\subsection{Eccentricity}

\begin{figure}
    \centering
    \includegraphics[width=.45\textwidth]{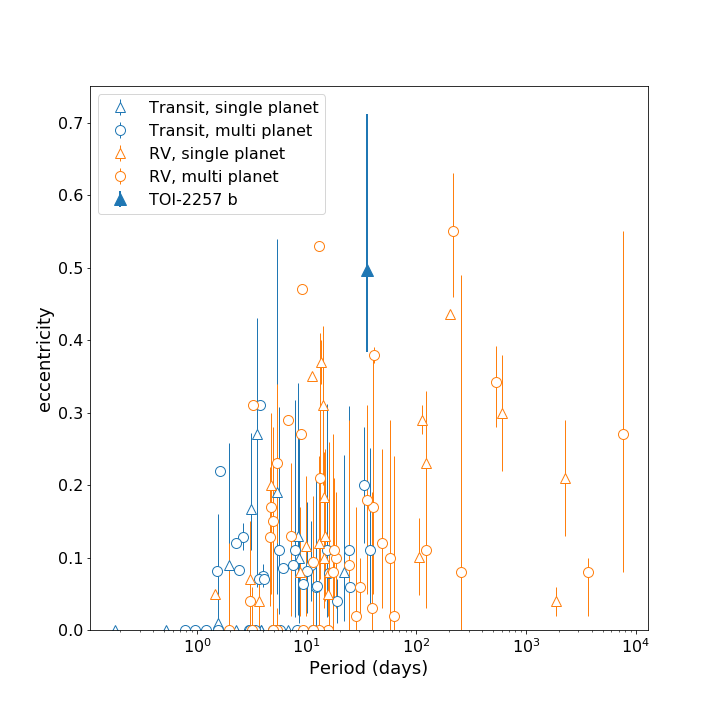}
    \caption{Period and eccentricities for known planets orbiting M-dwarf stars ($T_{\text{eff}} < 3700$). TOI-2257\,b, shown with a dark marker and thick errorbar for emphasis, is the most eccentric transiting planet with only 1 known planet in the system. Data retrieved from Exoplanet Archive\protect\footnotemark; note that systems without reported eccentricity errors are shown without errorbars.}
    
    \label{fig:period_eccen}
\end{figure}

Without radial velocity measurements, the eccentricity reported here is derived from the transit itself. In particular, it is driven by the transit duration, which can be degenerate with the impact parameter. Nevertheless, as the eccentric case is statistically favored over the circular case, we explore the system using the resulting eccentricity of $e=0.496_{-0.133}^{+0.216}$.
The high eccentricity of this object, shown in context in Figure \ref{fig:period_eccen}, provides a clue to the dynamic history of the system. Single-planet systems tend to have dynamically hotter orbits with larger eccentricities than those in multi-planet systems \citep{Xie2016,VanEylen2019,Masuda2020}. The high eccentricity of nearly 0.5 seen in TOI-2257\,b is suggestive of the possible influence of a long-period giant planet rather than self-excitation, the latter of which \cite{VanEylen2019} found to only explain eccentricities up to around 0.3. An example of this mechanism was found for GJ~1148\,b, a Saturn-mass planet orbiting an M dwarf in an eccentric orbit of $\sim$0.38 \citep{haghi2010}. Posterior observations found out that an outer giant planet was responsible for exciting GJ~1148\,b \citep{trifonov2018}.  
TOI-2257\,b's system parameters are consistent with the paradigm presented by \cite{Huang2017}, in which systems that survive scattering by an outer giant planet show lower multiplicity and higher eccentricity. Future radial velocity measurements would help to test this interpretation by searching for a massive planet in a more distant and likely non-transiting orbit. 

\footnotetext{\url{https://exoplanetarchive.ipac.caltech.edu/}}

\subsection{Prospects for Radial Velocity observations}

In order to constrain the mass of TOI-2257b, high-resolution spectroscopy with a precision of ${\sim}1$\,m\,s$^{-1}$ or less is needed. \cite{Demangeon21} showed in the case of the M3V-star L\,98-59, the mass of a planet with 2\,m\,s$^{-1}$ semi-amplitude can be constrained with about 10\% precision, taking the intrinsic stellar RV jitter into account. 
TOI-2257b is >3 mag fainter than L\,98-59, thus the RV precision is reduced to about 6-10\,m\,s$^{-1}$ for a 900\,s exposure with a state-of-the art instrument like ESPRESSO at the VLT \citep{pepe21}. Current and upcoming infrared spectrographs like CRIRES \citep{dorn14}, SPIROU \citep{donati20}, or NIRPS \citep{wildi17} will be able to obtain a higher signal-to-noise, but the RV-precision will still not allow direct measurement of the reflex motion of TOI-2257. However, the newly commissioned MAROON-X instrument \citep{MAROON-X2021} at the Gemini-North telescope is expected to achieve a signal of ~1\,m\,s$^{-1}$ for TOI-2257. Combining the instrument noise with an estimated stellar jitter of 1.5\,m\,s$^{-1}$, it would be possible to constrain the mass within 6\,$\sigma$ after 26 measurements with the red arm and exposure time of 20 minutes.

\begin{figure}[ht!]
    \centering
    \includegraphics[width=0.9\columnwidth]{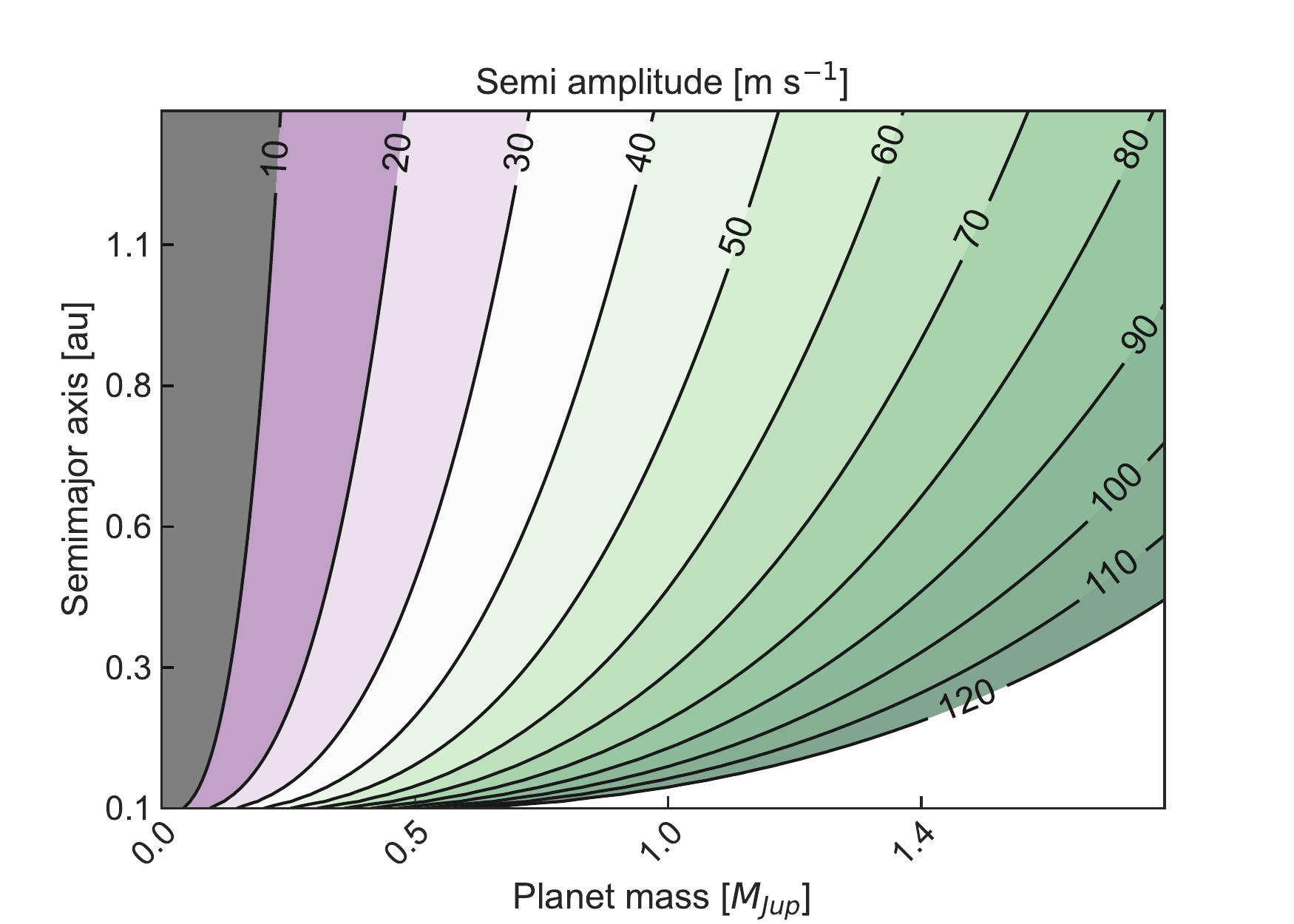}
    \caption{Expected Semi-amplitude of a possible giant planet orbiting TOI-2257 for different orbits and planetary masses.}
    \label{fig:TOI2257C_exp}
\end{figure}

Even without high-precision instruments, lower precision RV follow-up might result in the detection of a more massive planet, which is not transiting or further out and, thus, not visible in the current photometric data-set. We explore this possible case, since the presence of such a planet might cause the eccentricity of TOI-2257\,b. As shown in Fig.\,\ref{fig:TOI2257C_exp} a giant planet with 1\,$\rm M_{Jup}$, orbiting in 1\,au would result in a RV signal of about 50\,m\,s$^{-1}$ with an orbital period of 640\,days. The magenta and green areas in Fig.\,\ref{fig:TOI2257C_exp} show that the detection of an outer massive planet will be challenging, but possible for current high-resolution instruments. The expected long periods of such planets would require continuous monitoring over the course of months to several years with high-precision spectrographs.

\subsection{Prospects for atmospheric characterization}
Planet b's long orbit (35.9\,days) makes it a rare object among currently known sub-Neptunes, and thus a compelling target to study its atmospheric composition and gain clues about its formation and evolution.
The planet's low time-averaged equilibrium temperature\citep{Mendez2017} of $T_{\text{eq}}=256^{+61}_{-17}$\,K (assuming zero albedo and full day-night heat redistribution) along with the star's relative proximity to Earth (57.79\,pc) and optical/near-infrared brightness (K-mag 10.67) strongly favor transmission spectroscopy.
For example, JWST's transmission spectroscopy metric (TSM) is 32.71 following \citet{Kempton2018}.

Thus, TOI-2257\,b could be an interesting target to study ocean loss on this class of temperate sub-Neptunes. Probing O$_3$ abundances with red-sensitive JWST observations as well as O$_2$ abundances with ground-based optical observations from the Extremely Large Telescopes (e.g., \citealt{Luger2015, Serindag2019}), or probing CO and O$_4$ features \citep{Schwieterman2016}, could give clues in this regard.

\begin{figure}[ht!]
    \centering
    \includegraphics[width=0.9\columnwidth]{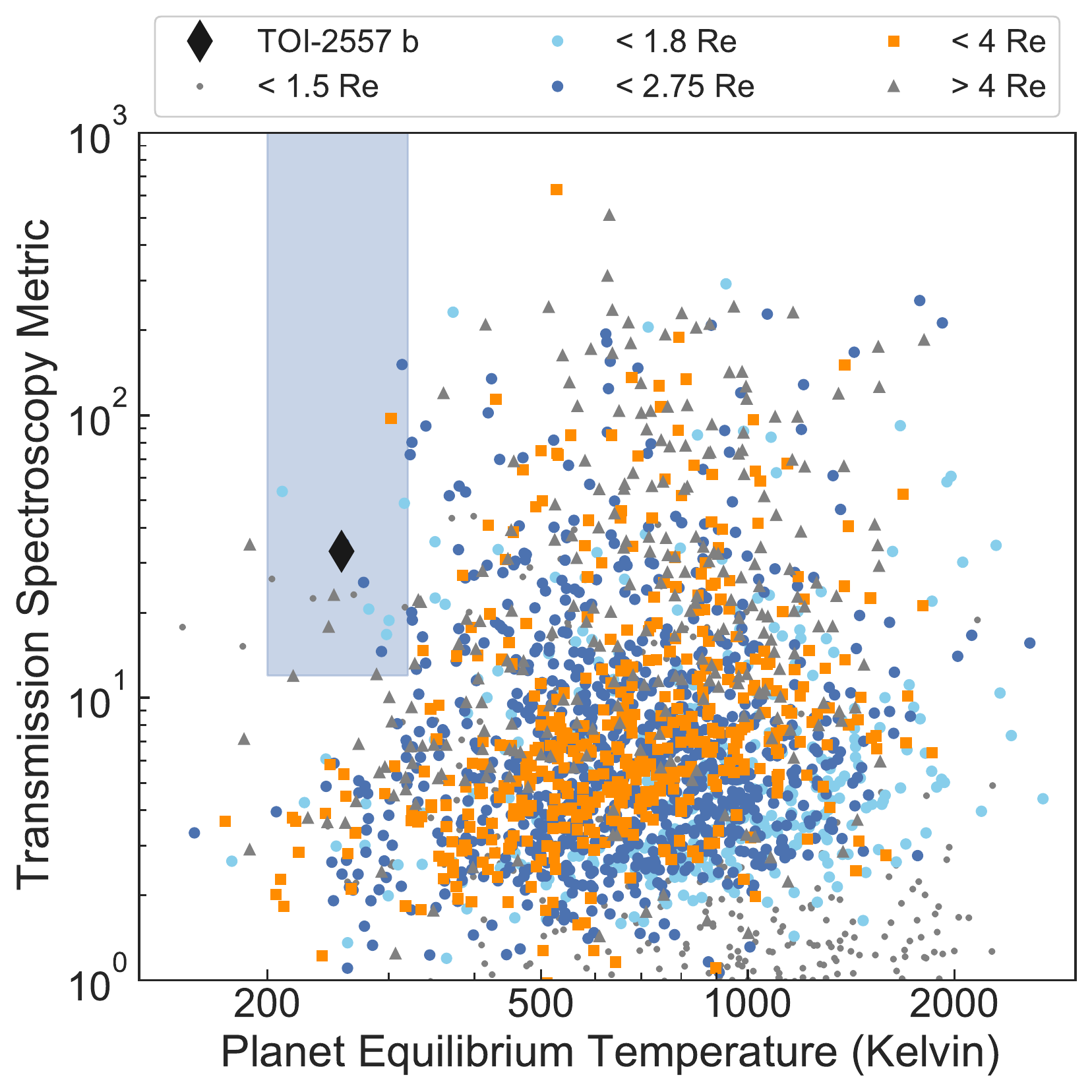}
    \caption{TOI-2257\,b and known exoplanets in the context of their equilibrium temperature (x-axis) and transmission spectroscopy metric \citep[TSM,][y-axis]{Kempton2018}. Different symbols/colors designate different planet size categories. TOI-2257\,b lies in the middle of the blue shaded area, which highlights sub-Neptunes that are (i) amenable for atmospheric characterization (TSM$>$12) and (ii) for which life in the clouds is possible ($T_{\text{eq}}$ between 200 and 320 K, following \citealt{Seager2021}). Equilibrium temperatures were estimated using an albedo of 0 and emissivity of 1, following \citet{Kempton2018}. Altering the assumptions for albedo and emissivity (e.g., assuming an albedo of 0.3 like for Earth and Neptune) adds lower error bars of $\sim$10\% and upper error bars of $\sim$30\% to the shown temperatures.
    Data retrieved from Exoplanet Archive\protect\footnotemark{} on 2021-07-30. Figure adapted from \citet{Seager2021}.}
    \label{fig:TSM}
\end{figure}

Finally, TOI-2257\,b is one of only $\sim$20 currently known, characterizable sub-Neptunes whose equilibrium temperatures fall into a liquid-water-is-possible zone (Fig.~\ref{fig:TSM}). Of course, such exoplanets do not allow surface habitability, given their immense atmospheric pressure and heated surface. However, life in the clouds could be a possibility--- as first proposed by \citep{Morowitz1967, Sagan1976} and recently discussed for sub-Neptune-sized exoplanets by \citet{Seager2021} (see also Fig.~\ref{fig:TSM}). 
Additionally, this temperature regime could favor the surface habitability conditions of any potential exomoons, as discussed in the following section.

\footnotetext{\url{https://exoplanetarchive.ipac.caltech.edu/}}


\subsection{Prospects of exomoons orbiting TOI-2257\,b}

Moons in the Solar System are invaluable puzzle pieces for understanding our system formation history, evolution, and even habitability \cite[see, e.g.,][]{opik1960,laskar1993,rufu2017,orgel2018}. 
Equally valuable is studying moons hosted in other planetary systems; that is, exomoons. Indeed, due to the large number of moons orbiting Solar System planets, the possible existence of moons orbiting exoplanets is widely accepted. However, despite efforts to find them \cite[see, e.g.,][]{kipping2012,kipping2015,hippke2015,Teachey2018,kreidberg2019} at the time of writing, no detection has yet to be confirmed. 

One of the major interests of exomoons is their potential habitability. Indeed, massive rocky exomoons orbiting temperate giant planets have been suggested as places where life might arise and evolve over long-time scales \cite[see, e.g.,][]{williams1997,kaltenegger2010,hellerbarnes2013,heller2014,hellerbarnes2015}. 
In particular, temperate moon--planet systems hosted by low-mass stars are widely discussed in the literature, providing arguments in favor and against the existence and habitability of exomoons. For example, it has been suggested that such moons will be likely tidally locked to the planet, a situation that would favor a uniform distribution of irradiation \citep{trifonov2020}. In addition, the magnetic field generated by the planet may protect the moon from the typical stellar flares produced by M dwarfs \citep{hellerzuluaga2013}. These factors would mitigate some typical inconveniences when discussing the habitability of a planet orbiting in the habitable zone of an M dwarf \citep{khoda2007,lammer2007}. On the other hand, gravitational perturbations by the close star and extra planets in the system could induce eccentricities that likely make any moon uninhabitable \citep{heller2012}. In any case, due to the number of unknown parameters to date, the field of habitable exomoons is still very speculative.

In this study, we focus only on estimating the existence of rocky exomoons orbiting TOI-2257\,b. To this end, we use dynamical considerations to 
compute the moon survival rate (MSR) following the formulation introduced by \cite{sasaki2012} and the prescriptions given by \cite{dobos2021}. The simulation details are provided in Appendix \ref{app:exomoon_simulation}. The mathematical description takes into account tidal evolution in a star--planet--moon system. It is important to notice that this formulation assumes that the planet resides in a circular orbit around the host star; however, this is not the case for TOI-2257\,b. Hence, the results yielded by our simulations should be considered as upper limits to the actual MSR, as eccentric planetary orbits reduce the stability of the moon--planet systems.

\begin{figure}
    \centering
    \includegraphics[width=.45\textwidth]{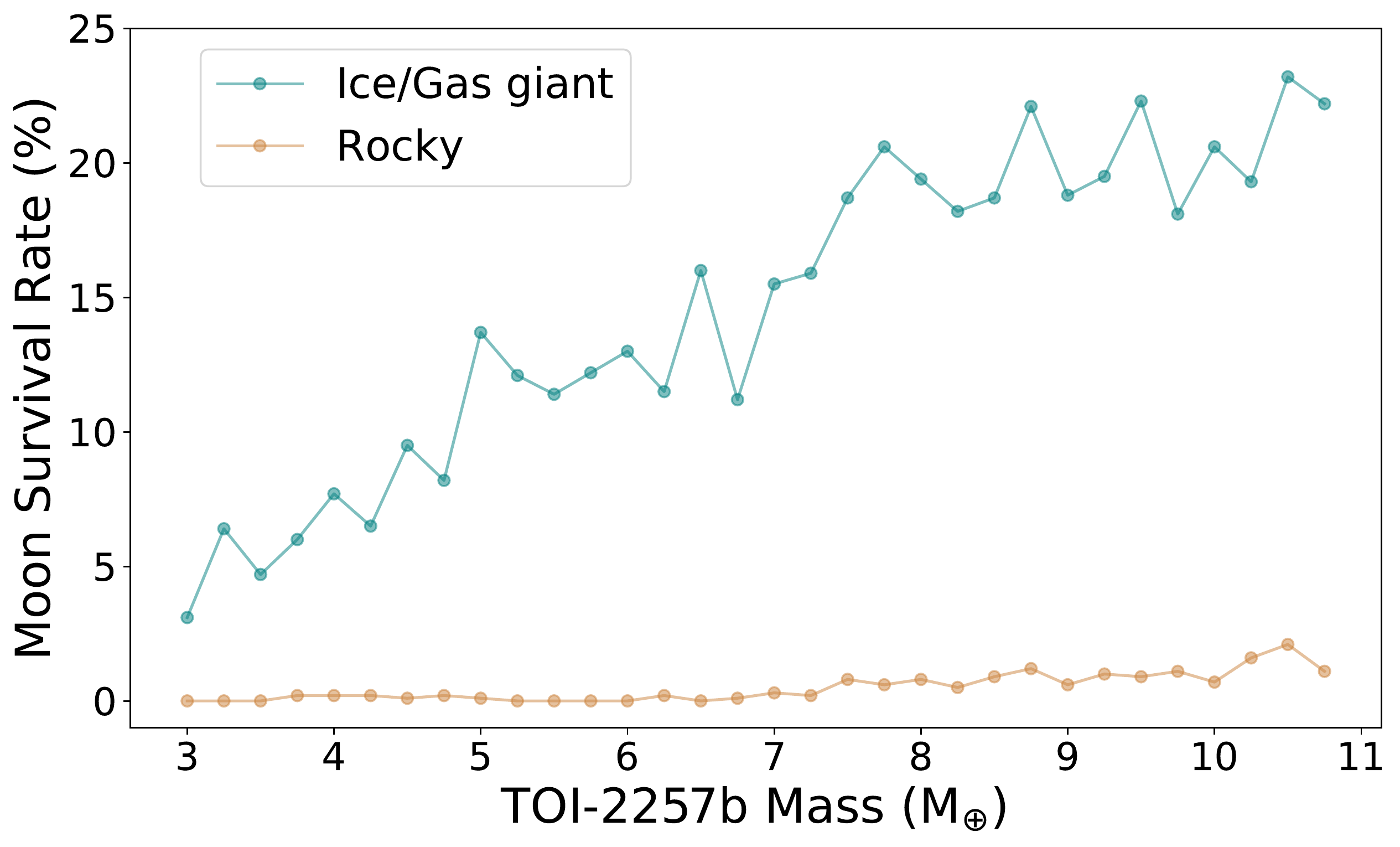}
    \caption{MSR upper limit around TOI-2257\,b as a function of the planetary mass. The blue line represents the solution when considering TOI-2257\,b as an ice/gas giant planet, while the yellow line is considering it a rocky planet. 
     Each data point is the MSR computed by simulating 1000 exomoons orbiting around TOI-2257\,b, for a total of 64,000 scenarios tested.}
    \label{fig:msr}
\end{figure}

The results of the simulations are displayed in Fig.~\ref{fig:msr}. In summary, we found that the larger the mass of the planet, the larger the MSR, with a maximum MSR of about 20\% for ice/gas giant planets. On the other hand, the MSR was almost null for the entire sample of possible planetary masses when considering a rocky planet. In reality, when increasing the planetary mass, the density also increases, passing from an ice/gas giant to a rocky planet. This change would happen between 5.5 and 6.5~M$_{\oplus}$, which combined with the planetary radius of 2.194~R$_{\oplus}$, yields a planetary density of 2.9--3.5~g~cm$^{-3}$. The MSR would follow the ice/gas giant solution until $\sim$6~M$_{\oplus}$, dropping then to the estimated MSR for rocky planets. In such a case, the maximum MSR, given for $\sim$6~M$_{\oplus}$, is about 13$\%$. 

We conclude that TOI-2257\,b is likely a single planet with a low probability of having any orbiting moons. This result is in line with those found by \cite{sasaki2012} and \cite{dobos2021}, who established that it is very challenging for planets orbiting low-mass stars with a semi-major axis $\leq$0.2~au to harbor moons in stable orbits.

\section{Conclusion} \label{sec:conclusion}
This work presents the discovery and characterization of TOI-2257\,b, a sub-Neptune in a relatively long-period orbit around an M3V star. The preliminary characterization is based on photometry from TESS and several ground-based facilities, with spectral analysis and high-resolution imaging supporting the validation. 
TOI-2257\,b occupies a sparsely-populated region of parameter space. Currently, only two other transiting exoplanets around M dwarfs with periods greater than that of TOI-2257\,b are known (Kepler-1652\,b \citep{Torres2017} and TOI-700\,d \citep{gilbert2020, rodriguez2020}). Furthermore, this single-planet system is the most eccentric planet transiting an M dwarf to date, providing an opportunity to test possible formation scenarios. 
 
While the expected reflex motion driven, on the order of a few m\,s$^{-1}$, make current observations challenging, recent and upcoming instrumentation at observatories with larger collecting area like MAROON-X at the Gemini Observatory and the HIRES spectrograph at the ELT \citep{Marconi21} will make it possible to obtain spectra for TOI-2257 with more than 20\,times higher SNR than with ESPRESSO. Thus, these large telescopes will enable us to derive the mass of TOI-2257\,b within 10\% and even allow us to search for possible other planets within the system. Regardless, this planet is one of only a small number of sub-Neptunes for which liquid water is a possibility. The expected TSM of the object make it amenable to more detailed atmospheric characterization from JWST.


\begin{acknowledgements}

N.S., R.W. and B.-O. D. acknowledge support from the Swiss National Science Foundation (PP00P2-163967 and PP00P2-190080). MNG acknowledges support from MIT's Kavli Institute as a Juan Carlos Torres Fellow and from the European Space Agency (ESA) as an ESA Research Fellow. A.A.B., B.S.S.and I.A.S. acknowledge the support of the Ministry of Science and Higher Education of the Russian Federation under the grant 075-15-2020-780 (N13.1902.21.0039). L.D. is an F.R.S.-FNRS Postdoctoral Researcher.
BVR thanks the Heising-Simons Foundation for support. This publication benefits from the support of the French Community of Belgium in the context of the FRIA Doctoral Grant awarded to M.T.
E. J. acknowledges DGAPA for his postdoctoral fellowship.
YGMC acknowledges support from UNAM-DGAPA PAPIIT BG-101321. 
D. D. acknowledges support from the TESS Guest Investigator Program grant 80NSSC19K1727 and NASA Exoplanet Research Program grant 18-2XRP18\_2-0136.
We acknowledge support from the Centre for Space and Habitability (CSH) of the University of Bern.
Part of this work received support from the National Centre for Competence in Research PlanetS, supported by the Swiss National Science Foundation (SNSF). 
Funding for the TESS mission is provided by NASA's Science Mission Directorate. We acknowledge the use of public TESS data from pipelines at the TESS Science Office and at the TESS Science Processing Operations Center. This research has made use of the Exoplanet Follow-up Observation Program website, which is operated by the California Institute of Technology, under contract with the National Aeronautics and Space Administration under the Exoplanet Exploration Program. Resources supporting this work were provided by the NASA High-End Computing (HEC) Program through the NASA Advanced Supercomputing (NAS) Division at Ames Research Center for the production of the SPOC data products. This paper includes data collected by the TESS mission that are publicly available from the Mikulski Archive for Space Telescopes (MAST).
This work is based upon observations carried out at the Observatorio Astron\'omico Nacional on the Sierra de San Pedro M\'artir (OAN-SPM), Baja California, M\'exico. We warmly thank the entire technical staff of the Observatorio Astron\'omico Nacional at San Pedro M\'artir in M\'exico for their unfailing support to SAINT-EX operations, namely:
E.~Cadena, T.~Calvario, E.~Colorado, F.~D\'iaz,  A.~Franco, B.~Garc\'ia, C.~Guerrero, G.~Guisa, F.~Guillen, A.~Landa, L.~Figueroa, B.~Hern\'andez, J.~Herrera, E.~L\'opez, E.~Lugo, B.~Mart\'inez, G.~Melgoza, F.~Montalvo, J.M.~Nu\~nez, J.L.~Ochoa, I.~Plauchu, F.~Quiroz, H.~Riesgo, H.~Serrano, T.~Verdugo, I. Zavala. 
The research leading to these results has received funding from the European Research Council (ERC) under the FP/2007--2013 ERC grant agreement n$^{\circ}$ 336480, and under the European Union's Horizon 2020 research and innovation programme (grants agreements n$^{\circ}$ 679030 \& 803193/BEBOP); from an Actions de Recherche Concert\'{e}e (ARC) grant, financed by the Wallonia--Brussels Federation, from the Balzan Prize Foundation, from the BELSPO/BRAIN2.0 research program (PORTAL project), from the Science and Technology Facilities Council (STFC; grant n$^\circ$ ST/S00193X/1), and from F.R.S-FNRS (Research Project ID T010920F). This work was also partially supported by a grant from the Simons Foundation (PI Queloz, grant number 327127), as well as by the MERAC foundation (PI Triaud). PI Gillon is F.R.S.-FNRS Senior Research Associate.
TRAPPIST is funded by the Belgian Fund for Scientific Research (Fond National de la Recherche Scientifique, FNRS) under the grant  PDR T.0120.21, with the participation of the Swiss National Science Fundation (SNF). MG and EJ are F.R.S.-FNRS Senior Research Associate.
This work makes use of observations from the LCOGT network. Part of the LCOGT telescope time was granted by NOIRLab through the Mid-Scale Innovations Program (MSIP). MSIP is funded by NSF.
Some of the observations in the paper made use of the High-Resolution Imaging instrument(s) ‘Alopeke (and/or Zorro). ‘Alopeke (and/or Zorro) was funded by the NASA Exoplanet Exploration Program and built at the NASA Ames Research Center by Steve B. Howell, Nic Scott, Elliott P. Horch, and Emmett Quigley. Data were reduced using a software pipeline originally written by Elliott Horch and Mark Everett. ‘Alopeke (and/or Zorro) was mounted on the Gemini North (and/or South) telescope of the international Gemini Observatory, a program of NSF’s OIR Lab, which is managed by the Association of Universities for Research in Astronomy (AURA) under a cooperative agreement with the National Science Foundation. on behalf of the Gemini partnership: the National Science Foundation (United States), National Research Council (Canada), Agencia Nacional de Investigación y Desarrollo (Chile), Ministerio de Ciencia, Tecnología e Innovación (Argentina), Ministério da Ciência, Tecnologia, Inovações e Comunicações (Brazil), and Korea Astronomy and Space Science Institute (Republic of Korea)
This research has made use of the NASA Exoplanet Archive, which is operated by the California Institute of Technology, under contract with the National Aeronautics and Space Administration under the Exoplanet Exploration Program.
This research made use of \textsf{exoplanet} \citep{exoplanet:joss, exoplanet:zenodo} and its dependencies \citep{exoplanet:agol20, exoplanet:arviz, exoplanet:astropy13, exoplanet:astropy18, exoplanet:kipping13, exoplanet:luger18, exoplanet:pymc3, exoplanet:theano}. Additional use of software packages AstroImageJ \citep{Collins2017} and TAPIR \citep{Jensen2013}.

\end{acknowledgements}


\bibliographystyle{aa} 
\bibliography{ref}

\onecolumn
\begin{appendix}

\section{Posterior distribution of transit parameters} \label{app:transit_posterior}

\begin{figure*}[h!]
    \centering
    \includegraphics[width=0.85\textwidth]{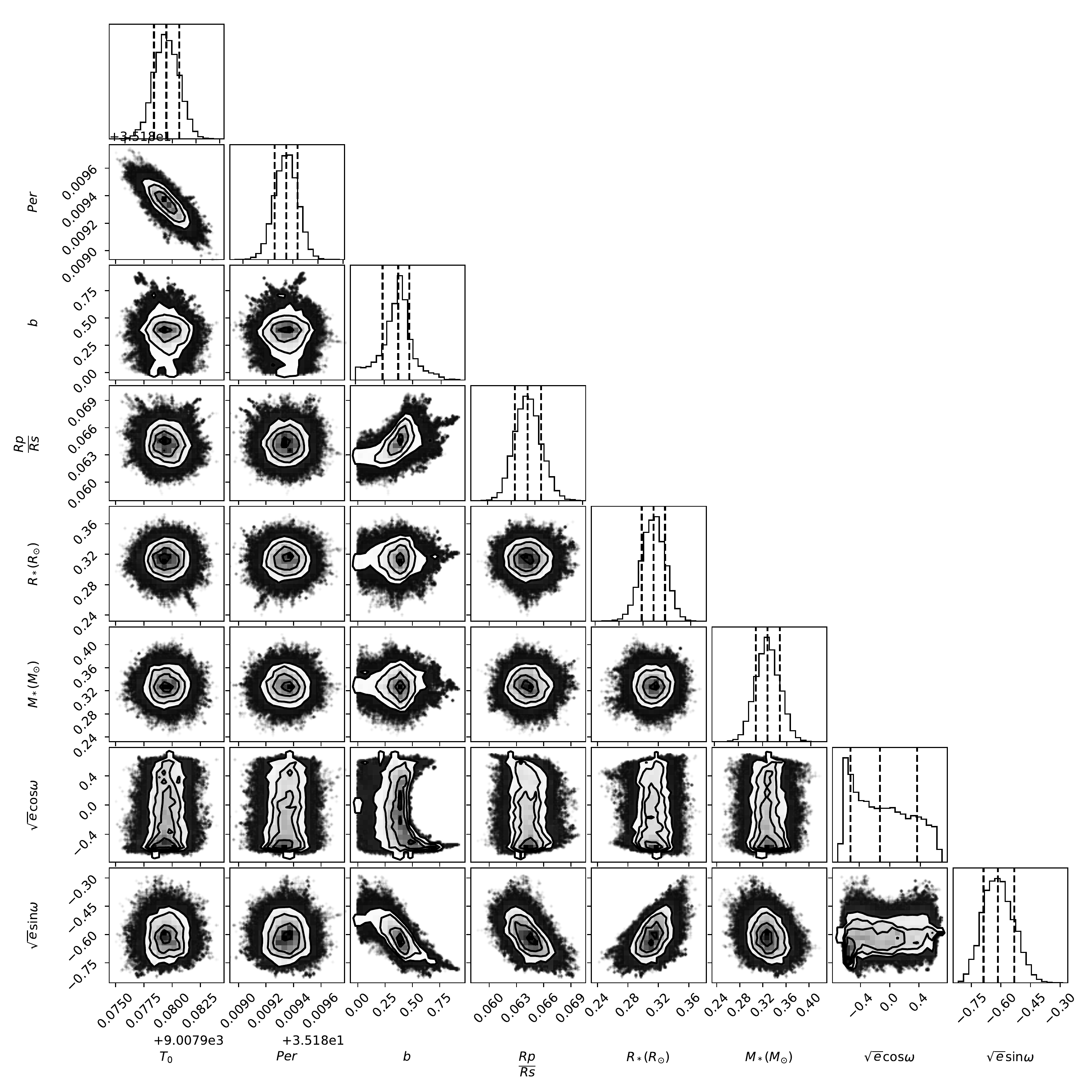}
    \caption{MCMC posterior distributions for the fitted transit parameters. From left-to-right: transit epoch (BJD$-$2,450,000), orbital period (d), impact parameter, planet-to-star radius ratio, stellar radius ($R_\ast$), stellar mass ($M_\ast$), $\mathrm{\sqrt{e}\cos\omega}$, and $\mathrm{\sqrt{e}\sin\omega}$. }
    \label{fig:mcmc}
\end{figure*}

\clearpage
\twocolumn
\section{Exomoon Simulations}\label{app:exomoon_simulation}
The main parameters that describe the stability of a moon are the planetary mass, radius and semi-major axis, the quality factor that describes the dissipation energy $Q_{p}$, the rotational state, and the potential Love number $k_{2p}$. 

In our case, only a few of these parameters are known: the planetary radius R$_{p}$=2.194~R$_{\oplus}$ and semi-major axis a$_{p}$=0.145~au. The other parameters are unknown. Hence, we explored them by taking into account the considerations described in detail by \cite{dobos2021}:
\begin{itemize}
    \item We explored the planetary mass M$_{p}$ considering its uncertainty given in Table~\ref{table:planet_params}; that is, ranging from 3 to 10.75~M$_{\oplus}$ in steps of 0.25~M$_{\oplus}$.
    \item The quality factor highly depends on the physical characteristics of the planet. It was established that for R$_{p}<2.0$~R$_{\oplus}$ (rocky planets), 10<Q$_{p}$<500, and for R$_{p}\geqslant2.0$~R$_{\oplus}$ (ice/gas giant planets) with orbital periods larger than 10~d, 10$^{3}$<Q$_{p}$<10$^{6}$, with the most probable value defined as $Q_{p}=3\times10^{4}$. Our estimation of the planetary radius is R$_{p}$=$2.194^{+0.113}_{-0.111}$~R$_{\oplus}$; that is, just in the limit between the given definition of rocky and ice/gas giant planets. To avoid as many biases as possible in our study, we ran two suites of simulations: one considering a rocky planet with $Q_{p}=2.5\times10^{2}$ and the other considering an ice/gas giant with $Q_{p}=3\times10^{4}$.
    \item The rotational spin of the planet was randomly explored between 10~h and 5~days.
    \item The potential Love number was assumed to be 0.299 for 
rocky planets and 0.5 for ice/gas giant planets. 
\end{itemize}

For each scenario, we generated a sample of 1000 moons whose stabilities were measured over a time--scale of 1~Gyr. The moons' parameters explored were: 
\begin{enumerate}
    \item The moon's mass, which was sampled randomly over the range (0.1--0.01)$\times$M$_{p}$ following a uniform distribution. 
    \item The moon's density, which was explored following a Gaussian distribution with a mean value of $\rho_{m}$=3~g~cm$^{-3}$ and $\sigma=1/3$~g~cm$^{-3}$.
    \item The moon's semi-major axis, a$_{m}$, which was chosen randomly from a uniform distribution between two times the Roche limit (R$_{limit}$) and the critical distance (C$_{d}$). 
\end{enumerate}

In each case, we monitored the dynamical evolution of the moon, prematurely finishing the integration if the moon collided with the planet or escaped from the system; that is, a$_{m}\leq$ R$_{limit}$ or a$_{m}\geq$ C$_{d}$, respectively. Then, the MSR was defined as the 
ratio of surviving moons (i.e. those that stayed in orbit around the planet until the end of the integration) and the number of tested configurations. In total, we explored 64,000 different scenarios. The results are explored further in the main text.



\end{appendix}

\end{document}